\documentclass[showpacs,preprintnumbers,amsmath,amssymb,twocolumn]{revtex4-1}
\usepackage{graphicx}
\usepackage{dcolumn}
\usepackage{bm}
\usepackage{subfigure}
\usepackage{float}
\usepackage{multirow}
\usepackage{booktabs}

\usepackage{amsmath}

\usepackage{latexsym,amssymb}
\usepackage[mathscr]{eucal}
\usepackage[switch*]{lineno}
\usepackage{cuted}
\usepackage[bookmarks=true,
   colorlinks=true,
   linkcolor=blue,
   urlcolor=blue,
   citecolor=blue,
   bookmarks=true,
   hyperindex=true
]{hyperref}

\begin{document}

\title{Phase transitions, critical behavior and microstructure of the FRW universe in the framework of higher order GUP}

\author{Zhong-Wen Feng\textsuperscript{1}}
\altaffiliation{Corresponding author: zwfengphy@cwnu.edu.cn}
\author{Shi-Yu Li\textsuperscript{1}}
\author{ Xia Zhou\textsuperscript{1}}
\author{Haximjan Abdusattar\textsuperscript{2}}
\altaffiliation{Corresponding author: axim@ksu.edu.cn}
\vskip 0.5cm
\affiliation{1 School of Physics and Astronomy, China West Normal University, Nanchong, 637009, China \\
2 School of Physics and Electrical Engineering, Kashi University, Kashi 844009, Xinjiang, China}

\begin{abstract}
In this paper, we explore the phase transition, critical behavior and microstructure of the FRW in the framework of a new higher order generalized uncertainty principle (GUP). Our initial step involves deriving the equation of state by defining the work density $W$ from GUP corrected Friedmann equations as the thermodynamic pressure $P$. Based on the modified equation of state, we conduct an analysis of the $P-V$ phase transition in the FRW universe. Subsequently, we obtain the critical exponents and coexistence curves for the small and large phases of the FRW universe around the critical point. Finally, employing Ruppeiner geometry, we derive the thermodynamic curvature scalar $R_N$, investigating its sign-changing curve and spinodal curve.  The results reveal distinctive thermodynamic properties for FRW universes with positive and negative GUP parameters $\beta$. In the case of $\beta>0$,  the phase transition, critical behavior and microstructure of FRW universe  are like those of Van der Waals system and charged AdS. Conversely, for $\beta<0$, the results resemble those obtained through effective scalar field theory. These findings underscore the capacity of quantum gravity to induce phase transitions in the universe, warranting further in-depth exploration.
\end{abstract}
\maketitle
\section{Introduction}
\label{intro}
As a pivotal concept in the realm of physics, thermodynamic phase transitions exert a profound influence on our comprehension of natural phenomena. These phase transition processes exist both at the microscopic atomic scale and throughout the macroscopic realm of astrophysics. Significant changes in energy, entropy, microstructure, and other properties in the transition of matter from one state to another have triggered an in-depth study of thermodynamic phase transitions. It is noteworthy that thermodynamic phase transitions extend beyond commonplace matter, manifesting also within gravitational systems. As an extreme manifestation of the gravitational field, the thermodynamics of black holes and their properties related to phase transitions have captivated the attention of both theoretical and astrophysical physicists \cite{ch1,ch2,ch3,ch3+,ch4}. Since the formulation of the four laws of black hole thermodynamics in the 1970s \cite{ch5}, considerable strides have been taken in unraveling the intricacies of black hole phase transitions \cite{ch6,ch7,ch8}. Subsequent to this, Mann et al. identified parallels between the phase transition behavior of AdS spacetime and that of Van der Waals (VdW) system, instigating a surge of interest in black hole physics \cite{ch8+,ch9,ch10,ch11}. Now, the thermodynamic phase transition of black holes can be used to further study their microstructure and interactions \cite{ch12,ch13,ch14,ch14b+,ch15b+,ch15,ch16,ch17}. Collectively, these findings underscore that thermodynamic phase transitions can serve as probes, facilitating a more nuanced analysis of gravity's characteristics.

On the other hand, the universe itself serves as an important gravitational system, whose properties expounded through the Friedmann-Robertson-Walker (FRW) metric. This widely embraced model encapsulates a well-defined horizon, a pivotal element supporting a self-consistent thermodynamic framework. Therefore, the universe establishes a profound physical interconnection with thermodynamics. In analogy to the rich thermodynamic phenomena observed in black holes, the FRW universe unfolds numerous intriguing aspects, including the Hawking temperature, Bekenstein-Hawking entropy, quasi-local energy, \emph{et al}. \cite{ch17+,ch18,ch19,ch20,ch20+,ch21,ch21+}. Notably, recent investigations into the thermodynamic properties of the classical FRW universe by Haximjan \emph{et al}. revealed the absence of a $P-V$ phase transition \cite{ch22}. However, a revelation emerges when shifting the focus to the realm of modified gravity. In Ref.~\cite{ch23}, Kong \emph{et al}. utilized the Horndeski gravity to modify the pressure and density of the FRW universe with a perfect fluid and successfully construct a VdW-like equations of state, which allow them to analyze the phase transitions and critical behavior therein. Interestingly, their findings diverged from conventional expectations, demonstrating that the coexisting phase of the $P-V$ transition materializes above a critical temperature-a departure from the anticipated behavior in typical VdW fluids and many black hole systems. Building on this groundwork, subsequent research has delved into the exploration of phase transitions, critical behavior, and microstructure in the FRW universe within the framework of various modified gravity theories~\cite{ch24,ch25,ch26,ch27,ch27+}.

The above works suggest that thermodynamic phase transitions may occur in the models of the FRW universe that are built beyond classical general relativity (GR). Up now, there are many schemes for modifying GR, among them the most notable one must be quantum gravity (QG). It is well known that the effects of QG  were relatively strong in the early universe, while astronomical observations have suggested a significant number of phase transition processes during that era~\cite{cha28,cha29,cha30,cha31}. This raises our curiosity about whether the models of QG that also go beyond GR, such as the generalized uncertainty principle (GUP), could instigate thermodynamic phase transitions in the universe?  Indeed, numerous studies have shown that the GUP can modify the properties of the FRW universe~\cite{ch28,ch29,ch29+,ch29++,ch30,ch32,ch34,ch35,ch36,ch37}. For example, in Ref.~\cite{ch29} the GUP corrected dynamics of the universe is investigated, the results showed that effect of GUP can avoid the Big Bang singularity. Moussa \emph{et al}.~\cite{ch30,ch32} used GUP to modify the properties of early universe and consequently analyze the stochastic gravitational wave therein, it was argued that this was a way to test the effectiveness of the GUP theory. Besides, the GUP can correct Friedmann equations and thereby change the pressure and density of the universe, providing a feasible solution to the baryon asymmetry. In Ref.~\cite{ch37}, the authors studied the impact of GUP and the laws of thermodynamics on Friedmann equations. These investigations reveal that the attributes of the FRW universe within the GUP framework are more intricate than those in the classical scenario. This raises the possibility that thermodynamic phase transitions might manifest in these scenarios, warranting further in-depth exploration. However, this issue has not yet been discussed.

Based on the above statements and considerations, we attempt to explore the influence of the GUP on the thermodynamic properties of the universe and discuss whether it can lead to a phase transition.  Notably, we do not use the two popular types of GUP models as most of the previous works, namely, KMM model ($\Delta x\Delta p \geq \hbar [1 + {\beta _0}\ell _p^2\Delta {p^2}/{\hbar ^2}]/2$) \cite{ch40} , and the ADV model ($\Delta x\Delta p \geq \hbar [1 - {\alpha _0}\ell _p^2\Delta p/\hbar  + {\beta _0}\ell _p^2\Delta {p^2}/{\hbar ^2}]/2$  with the GUP parameters $\beta _0$  and  $\alpha _0$) \cite{ch41} (the KMM model can be seen as a special case of the ADV model at   $\alpha _0=0$), since they have the following: first of all, the perturbations of these models are valid only for small values of the GUP parameter; second, they do not imply the non-commutative geometry; third, absence of the value of the observed momentum, this leads to the models are not applicable to double special relativity \cite{ch42}. In order to address these limitations, the higher-order GUP has been proposed. This non-perturbative model is consistent with various proposals for QG and does not contradict double special relativity, and has therefore received the attention \cite{ch43}. Now, much research is focused on the construction and application of higher-order GUP  \cite{ch44,ch45,ch46,ch47,ch48,ch49}.

In light of the aforementioned considerations, it can be seen that the higher-order GUP formulations harbor the potential to instigate phase transitions in the FRW universe. Such transitions offer a pathway for comprehending the thermodynamic intricacies in the universe. Recently, by considering the minimum length is a model-independent feature of QG,  a new higher-order GUP as follows~\cite{ch50} (In fact, the expression can be called the extended GUP according to the definitions provided in Refs.~\cite{ch50+,ch51,ch52,ch53} since it is no longer related to $\Delta p$  but to $\Delta x$. However, for the sake of consistency, we will continue to use the designation in the original literature)

\vspace{-\baselineskip}
\begin{align}
\label{eq1}
\Delta x\Delta p \geq \frac{\hbar }{2}\frac{1}{{1 + \left( {{{16\beta } \mathord{\left/ {\vphantom {{16\beta } {\Delta {x^2}}}} \right. \kern-\nulldelimiterspace} {\Delta {x^2}}}} \right)}},
\end{align}
where $\beta  = {\beta _0}\ell _p^2$  with the deformation parameter (or  GUP parameter) and  ${\ell _p}$ the Planck length. $\Delta x$  and  $\Delta p$ are the uncertainties for position and momentum, respectively. By analyzing Eq.~(\ref{eq1}), one can figure out that it has two main characteristics: (i) The deformation parameter   are no longer restricted to positive values, but negative values are also allowed. However, Eq.~(\ref{eq1}) recovers to the conventional Heisenberg uncertainty principle  $\Delta x\Delta p \geq {\hbar  \mathord{\left/ {\vphantom {\hbar  2}} \right. \kern-\nulldelimiterspace} 2}$ when  ${\beta _0} = 0$. (ii) It leads to a fixed and consistent minimum length for both cases positive and negative deformation parameters, which indicate that this new model has a better parameter adaptability for minimum length than other ones. The validity of the QG effect in the model is ensured by these advantages, which allow us to analyze the effects of both positive and negative deformation parameters on the same physical system. In this work, according to the new higher-order GUP~(\ref{eq1}), we first derive the modified Friedmann equations of FRW universe. Then, based on these modifications, the thermodynamic pressure and equation of state in the universe can be obtained, which allows us to further analyze the phase transitions and critical behavior therein. Finally, by utilizing Ruppeiner geometry, we discuss the microscopic properties of the FRW universe through the analysis of the behavior of the reduced thermodynamic curvature.

The outline of our paper is as follows. In Section~\ref{sec2}, we shortly review the new higher-order GUP corrected Friedmann equations, then we derive the  corresponding modified thermodynamic equation of state of FRW universe. In Section~\ref{sec3}, we analyze its phase transition, critical behaviors and the microstructure of the FRW universe in the framework of QG. The paper ends with conclusions in section~\ref{sec4}.

To simplify the notation, this research takes the  units $G=c={k_B}=1$.

\section{Modified Friedmann equations and the corresponding equation of state of FRW universe}
\label{sec2}
To begin with, we briefly review the new higher-order GUP corrected Friedmann equations,  and then derive the corresponding modified equation of state of FRW universe. To the best of our knowledge, there are two valid and feasible schemes for deriving the Friedmann equation. The first, referred to as the classical scheme, involves deriving the energy density and pressure of the FRW universe from the action of a gravity theory, thereby yielding the Friedmann equation. The second scheme is grounded in the assumption that the apparent horizon of the universe is connected to geometric entropy. By applying this entropy to the first law of thermodynamics, one can consequently derive the Friedmann equations \cite{ch18}. It is noteworthy that both schemes produced equivalent results. However, given the facile modification of thermodynamic quantities within the GUP framework, our previous work~\cite{ch54} leaned towards the second scheme. According to Eq.~(\ref{eq1}), we derived the GUP corrected geometric entropy  (a detailed derivation is presented in~\ref{appA}) as follows~\cite{ch54}

\vspace{-\baselineskip}
\begin{align}
\label{eq1+}
{S_{{\text{GUP}}}} = \frac{A}{4} + 4\pi \beta \ln \left( {\frac{A}{{{A_0}}}} \right),
\end{align}
where $A$ is the area of the horizon and  $A_0$ is a constant with units of area.  In the works on the phase transition of the universe~\cite{ch23,ch24,ch25}, it has been found that the entropy of the universe has a similar form of expression as the above equation, which indicates that it is also possible to obtain the P-V phase transition of the universe in the framework of the GUP based on Eq.~(\ref{eq1+}). Following the second scheme, the corresponding modified Friedmann equations (see a more detailed derivation in the~\ref{appB}) are given by~\cite{ch54}

\vspace{-\baselineskip}
\begin{subequations}
\label{eq2}
\begin{align}
 - 4\pi (\rho  + p) &= \dot H\left( {1 + 4\beta {H^2}} \right),
 \\
\frac{8}{3}\pi \rho & = {H^2}\left( {1 + 2\beta {H^2}} \right),
\end{align}
\end{subequations}
where  $H$,   $\rho$ and  $p$ represent the Hubble parameter, energy density and pressure of the universe, respectively. The dot stands for the derivative with respect to the cosmic time. According to Eq.~(\ref{eq2}), the GUP corrected energy density and pressure are be expressed as

\vspace{-\baselineskip}
\begin{subequations}
\label{eq3}
\begin{align}
\rho  & = \frac{{3{H^2}}}{{8\pi }} + \frac{{3{H^4}\beta }}{{4\pi }},
 \\
p & =  - \frac{{\dot H}}{{4\pi }} - \frac{{3{H^2}}}{{8\pi }} - \frac{{\dot H{H^2}\beta }}{\pi } - \frac{{3{H^4}\beta }}{{4\pi }}.
\end{align}
\end{subequations}
By using the condition  ${h^{a b }}\left( {{\partial _a}R} \right)\left( {{\partial _b }R} \right) = 0$, the apparent horizon of the FRW universe reads  $R_A = {1 \mathord{\left/ {\vphantom {1 H}} \right. \kern-\nulldelimiterspace} H}$, and its time derivative is  $\dot R_A \! = \!  - \dot H{R_A^2}$ \cite{ch19}. Therefore, one can rewritten Eq.~(\ref{eq3}) in term of $R_A$ and $\dot R_A$ as follows

\vspace{-\baselineskip}
\begin{subequations}
\label{eq5}
\begin{align}
\rho \left( {R_A,\dot R_A} \right) & = \frac{3}{{8\pi {R_A^2}}} + \frac{{3\beta }}{{4\pi {R_A^4}}},
 \\
p\left( {R_A,\dot R_A} \right) & =  - \frac{3}{{8\pi {R_A^2}}} + \frac{{\dot R_A}}{{4\pi {R_A^2}}} - \frac{{3\beta }}{{4\pi {R_A^4}}} + \frac{{\dot R_A \beta }}{{\pi {R_A^4}}}.
\end{align}
\end{subequations}
 Then,  taking account of the definition of the work density of a matter field  $W =  - {{{h_{a b }}{T^{a b }}} \mathord{\left/ {\vphantom {{{h_{a b}}{T^{a b}}} 2}} \right. \kern-\nulldelimiterspace} 2} = {{\left( {\rho  - p} \right)} \mathord{\left/ {\vphantom {{\left( {\rho  - p} \right)} 2}} \right.
 \kern-\nulldelimiterspace} 2}$ \cite{ch3,ch3+}, one has

 \vspace{-\baselineskip}
 \begin{align}
\label{eq6}
 W &= \frac{1}{2}\left[ {\rho \left( {R_A,\dot R_A} \right) - p\left( {R_A,\dot R_A} \right)} \right]
\nonumber \\
 & = \frac{3}{{8\pi {R_A^2}}} - \frac{{\dot R_A}}{{8\pi {R_A^2}}} + \frac{{3\beta }}{{4\pi {R_A^4}}} - \frac{{\dot R_A\beta }}{{2\pi {R_A^4}}},
 \end{align}
 and the Misner-Sharp energy is

\vspace{-\baselineskip}
\begin{align}
\label{eq6+}
E = \rho \left( {R_A,\dot R_A} \right)  V = \frac{{ {{R_A^2} + 2\beta }}}{{2R_A}},
 \end{align}
where  $V = {{4\pi {R_A^3}} \mathord{\left/ {\vphantom {{4\pi {R_A^3}} 3}} \right. \kern-\nulldelimiterspace} 3}$ is the volume of FRW universe.

Next, it is necessary to discuss the temperature of the system. In the FRW universe, the surface gravity on the apparent horizon can be defined as $\kappa  = {\partial _a }\left( {\sqrt { - h} {h^{a b }}{\partial _a }r} \right)$ ${\left. {/2\sqrt { - h} } \right|_{R = {R_A}}}$.  With the help of  $\kappa$, the temperature of  spatially flat FRW universe is given by

\vspace{-\baselineskip}
\begin{align}
\label{eq7}
T = \frac{{\left| \kappa  \right|}}{{2\pi }} = \frac{1}{{2\pi R_A}}\left( {1 - \frac{{\dot R_A}}{2}} \right).
\end{align}
It is important to note that the ${\dot R_A} \ll 1$ is a small quantity, which leads to FRW universe has an inner trapping horizon. Besides, in Ref.~\cite{ch18}, the authors pointed out it is not necessary to  consider the  change in temperature in the application of the first law of thermodynamics. Therefore, when employing the first law of thermodynamics and geometric entropy to derive the Friedmann equations, the small quantity ${\dot R_A}$ can be neglected, and the classical Hawking temperature $T = {1 \mathord{\left/ {\vphantom {1 {2\pi {R_A}}}} \right. \kern-\nulldelimiterspace} {2\pi {R_A}}}$ is reinstated. However, for further discussion of the phase transitions of the universe, it should be considered as undergoing slow evolution, meaning its thermodynamic processes are quasi-static.  Therefore, ${\dot R_A}$ is retained in the next study. Now, according to Eq.~(\ref{eq6})-Eq.~(\ref{eq7}),  it can be demonstrated that the above defined quantities are satisfy the thermodynamic first law  ${\text{d}}E \! = \!  - T{\text{d}}S + W{\text{d}}V$ (see~\ref{appC} for more details about the first law of thermodynamics)  \cite{ch20+}. When comparing it with  the standard form of the first law of thermodynamics ${\text{d}}U  \! =  \!  - T{\text{d}}S + P{\text{d}}V$, it is discerned that the Misner-Sharp energy should be interpreted as $E  \! \equiv   \! - U$, and more importantly, the thermodynamic pressure can be defined as  $P \equiv W$. It is evident that neglecting ${\dot R_A}$  would result in a density equal to zero, making it impossible to define the thermodynamic pressure. Therefore, the inclusion of ${\dot R_A}$ is important for the present study. Now, combining Eq.~(\ref{eq6}) and Eq.~(\ref{eq7}), the modified thermodynamic equation of state for the FRW universe is expressed as

\vspace{-\baselineskip}
\begin{align}
\label{eq8}
P & = T\left[ {{{\left( {\frac{\pi }{{6V}}} \right)}^{1/3}} + \frac{{8\pi \beta }}{{3V}}} \right]
\nonumber \\
&+ \frac{1}{{{V^{4/3}}}}\frac{{3 \times {6^{1/3}}{V^{2/3}} - 4 \times {{\left( {6\pi }\right)}^{2/3}}\beta }}{{36{\pi ^{1/3}}}}.
\end{align}
The above equation illustrates that the influence of QG alters the pressure by adjusting for the volume (or size) and temperature of the FRW universe, thereby creating conditions conducive to undergoing a phase transition. Nevertheless, when ignoring the GUP parameter $\beta$, Eq.~(\ref{eq8}) simplified to the one in Einstein gravity, which has no phase transition, see Ref.~\cite{ch23}.

\section{The phase transition, critical behaviors and microstructure of the FRW universe in the framework of GUP}
\label{sec3}
In this section, we are going to investigate the phase transitions, critical behaviors and microstructure of FRW universe based on the modified equation of state. Based on the above discussion, it can be seen that the FRW universe satisfies the first law of classical thermodynamics. Therefore, based on the Refs.~\cite{ch22,ch23,ch24,ch25,ch26,ch27,ch27+}, it it undergoes a thermodynamic phase transition when the following conditions are satisfied

\vspace{-\baselineskip}
\begin{subequations}
\label{eq9}
\begin{align}
{\left( {\frac{{\partial P}}{{\partial V}}} \right)_T} = {\left( {\frac{{{\partial ^2}P}}{{\partial {V^2}}}} \right)_T} = 0, \\
{\left( {\frac{{\partial P}}{{\partial R_A}}} \right)_T} = {\left( {\frac{{{\partial ^2}P}}{{\partial R_A^2}}} \right)_T} = 0.
\end{align}
\end{subequations}
Therefore, by substituting Eq.~(\ref{eq8}) into Eq.~(\ref{eq9}), one can obtain two nonnegative volumes as follows

\vspace{-\baselineskip}
\begin{subequations}
\label{eq10}
\begin{align}
{V_{c1}} = \frac{4}{3}\pi R_{c1}^3= \frac{{32}}{3}\pi {\left[ {\left( {3 + 2\sqrt 3 } \right)\beta } \right]^{3/2}},  \label{eq10-1}\\
{V_{c2}} = \frac{4}{3}\pi R_{c2}^3= \frac{{32}}{3}\pi {\left[ {\left( {3 - 2\sqrt 3 } \right)\beta } \right]^{3/2}} \label{eq10-2}.
\end{align}
\end{subequations}
To ensure the above two results are real,  it is necessary to constrain the range of $\beta$. Eq.~(\ref{eq10-1}) demands $\beta  > 0$, while Eq.~(\ref{eq10-2}) requires  $\beta  < 0$. Fortunately, the  GUP parameter of Eq.~(\ref{eq1}) is just satisfy these requirements. Meanwhile, it is also worth noting that positive and negative values of the GUP parameter generate different QG effects (see Ref.~\cite{ch50} for the details), potentially leading to different properties of the phase transition, criticality, and microstructure of the FRW universe. Consequently, we will discuss these two scenarios separately below

\subsection{The phase transition, critical behaviors and microstructure for $\beta  > 0$}
\label{sec3.1}
\subsubsection{The phase transition}
\label{sec3.1.1}
By using Eq.~(\ref{eq8})-Eq.~(\ref{eq10}), the critical radius of apparent horizon, temperature, and pressure for $\beta  > 0$ are given by

\vspace{-\baselineskip}
\begin{align}
\label{eq11}
{R_c}  & = 2\sqrt {\left( {3 + 2\sqrt 3 } \right)\beta }, {V_c}  = \frac{4}{3}\pi R_c^3,
\nonumber \\
{T_c} & =  - \frac{1}{{12\pi \sqrt {\left( {1 + \frac{2}{{\sqrt 3 }}} \right)\beta } }}, {P_c}  = \frac{{8\sqrt 3  - 15}}{{576\pi \beta }}.
\end{align}
One special property of FRW universe in the framework the new higher order GUP with $\beta  > 0$ is its critical temperature is negative(the similar outcomes have been observed in Ref.~\cite{ch58}). It should be noted that negative temperature is not a new  concept.  It was first proposed by Onsager \cite{chz1} and experimentally discovered by Purcell and Pound \cite{chz2} during their study of nuclear spin systems in LiF crystals. Then, Ramsey \cite{chz3} provided a theoretical explanation of this phenomenon in 1956. In the 21st century, with advancements in quantum mechanics and cold atom technology, the notion of negative temperature has been tested in a wider range of physical systems (see, e.g., Refs.\cite{chz4,chz5,chz6}). However, the fact that negative temperature violates the basic principles of classical thermodynamics, in particular the relationship of temperature to energy and entropy, and that a broader validation of the phenomenon of negative temperature has not yet been accomplished, renders the concept still controversial \cite{,chz7,chz8,chz9,ch54+}. Nevertheless, it is believed that the negative temperature is potentially linked to physical processes in extreme conditions in cosmology \cite{chz6,ch56+,ch56+2,ch56+3,ch56+4,ch56,ch57}. Therefore, it will be relevant to discuss the phase transition, critical behavior and microstructure of the universe at negative temperatures. The above critical quantities in turn yield a dimensionless constant

\vspace{-\baselineskip}
\begin{align}
\label{eq12}
\chi =\frac{{{\nu _c}{P_c}}}{{{T_c}}} = \frac{{6 - \sqrt 3 }}{{12}} \approx 0.355,
\end{align}
where $\nu _c=2R_c $ is the specific volume.  It is clear that Eq.~(\ref{eq12}) is slightly less than that of VdW system ${\chi _{{\text{VdW}}}} = {3 \mathord{\left/ {\vphantom {3 8}} \right. \kern-\nulldelimiterspace} 8} \approx 0.375$, which indicates the thermodynamic properties of FRW universe with positive GUP parameter like those in the previous studies~\cite{ch8+,ch9,ch10,ch11,ch12,ch13,ch14,ch15,ch16,ch17}. In order to confirm this conjecture, the reduced volume, temperature and pressure can be defined for convenience as follows \cite{ch24}

\vspace{-\baselineskip}
\begin{align}
\label{eq13}
\tilde V = \frac{V}{{{V_c}}},\tilde T = \frac{T}{{{T_c}}},\tilde P = \frac{P}{{{P_c}}},
\end{align}
Substituting Eq.~(\ref{eq13}) into Eq.~(\ref{eq8}), the modified reduced thermodynamic equation of state is rewritten as

\begin{figure}[htbp]
\centering
\includegraphics[width=0.32\textwidth]{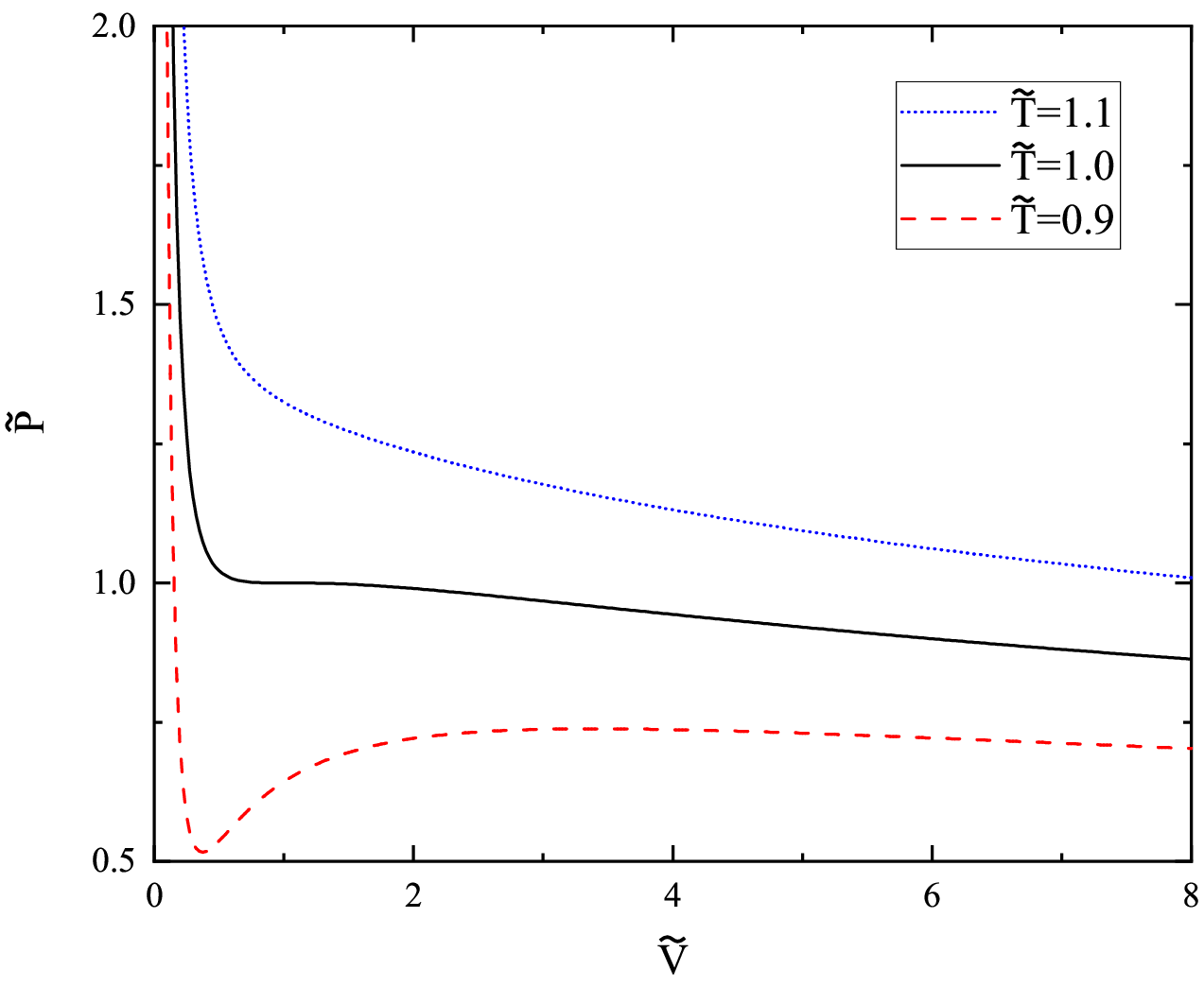}
\caption{\label{fig1} Variation in reduced pressure with reduced volume for different reduced temperature with $\beta>0$.}
\end{figure}

\vspace{-\baselineskip}
\begin{align}
\label{eq14}
\tilde P &= \frac{{3\left( {7 + 4\sqrt 3 } \right)}}{{\left( {111 + 64\sqrt 3 } \right){{\tilde V}^{4/3}}}} + \frac{{4\left( {12 + 7\sqrt 3 } \right)\tilde T}}{{\left( {111 + 64\sqrt 3 } \right)\tilde V}}
\nonumber \\
&- \frac{{6\left( {45 + 26\sqrt 3 } \right)}}{{\left( {111 + 64\sqrt 3 } \right){{\tilde V}^{2/3}}}} + \frac{{12\left( {26 + 15\sqrt 3 } \right)\tilde T}}{{\left( {111 + 64\sqrt 3 } \right){{\tilde V}^{1/3}}}}.
\end{align}
According to Eq.~(\ref{eq14}), the reduced pressure as a function of reduced volume for different reduced temperature is depicted in Fig.~\ref{fig1}.

In Fig.~\ref{fig1}, we depict the isothermal curves in the $P-V$ plane, with the reduced temperature of isotherms  $\tilde T$ decreasing from top to bottom. The red curve corresponds to   $\tilde T>1$ (or  $T > {T_c}$) and exhibits an ideal gas-like one-phase behavior, indicating the stability of the FRW universe without undergoing a phase transition. These observed behaviors bear similarity to those of VdW gas, implying a profound connection between the new higher-order GUP with a positive deformation parameter and VdW gas.

The same conclusion can also be inferred from the relationship between the Gibbs free energy and pressure. By using the usual thermodynamic relations \cite{ch27}

\begin{figure}[htbp]
\centering
\includegraphics[width=0.32\textwidth]{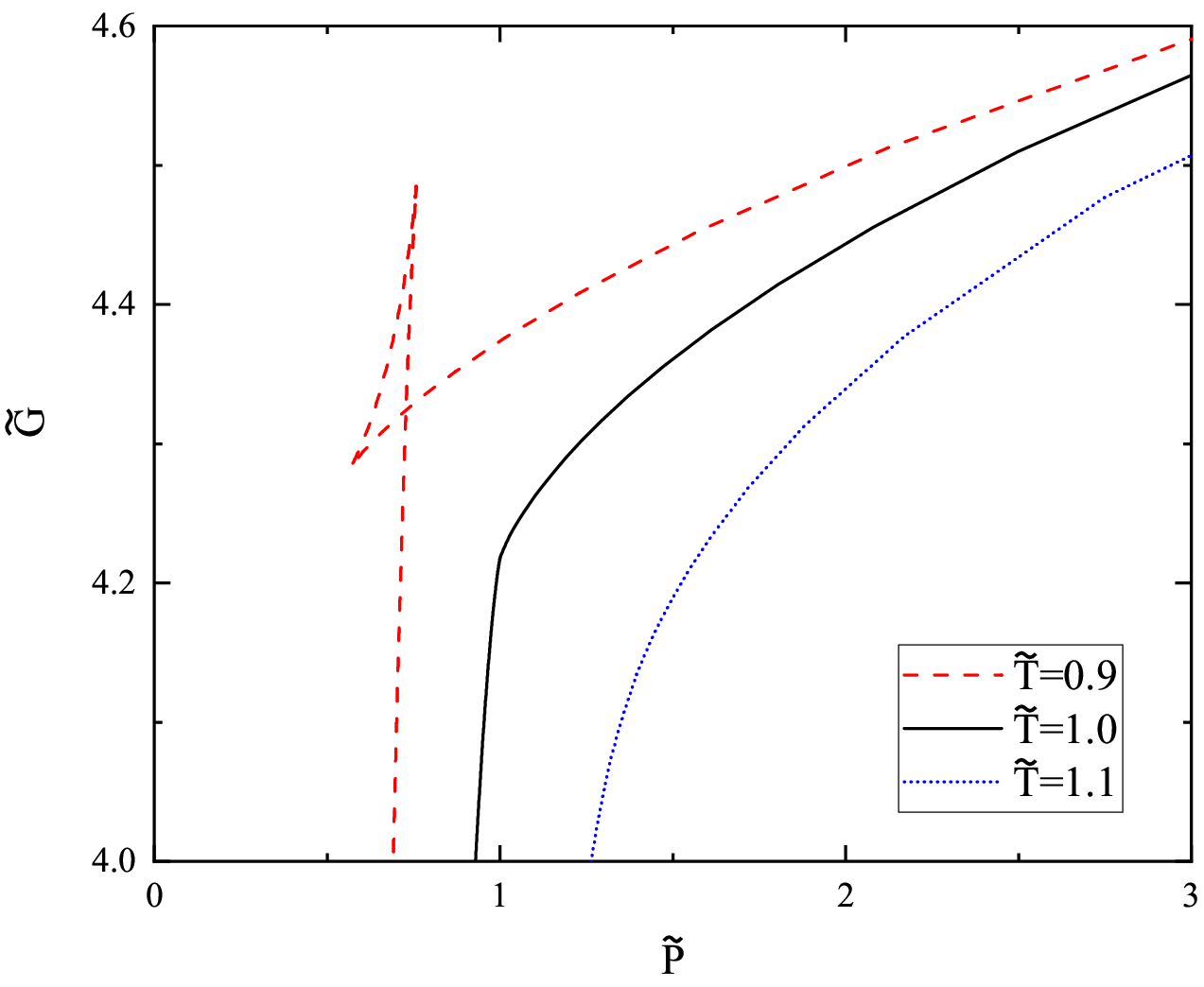}
\caption{\label{fig2} Variation in reduced Gibbs free energy with reduced pressure for different reduced temperature with $\beta>0$.}
\end{figure}

\vspace{-\baselineskip}
\begin{align}
\label{eq15+}
{\left. {\left( {\frac{{\partial \tilde G}}{{\partial \tilde V}}} \right)} \right|_{\tilde T}} = \tilde V{\left. {\left( {\frac{{\partial \tilde P}}{{\partial \tilde V}}} \right)} \right|_{\tilde T}},
\end{align}
one can obtain the reduced Gibbs free energy $\tilde G$. From Eq.~(\ref{eq15+}), we plot the behavior of the reduced Gibbs free energy  $\tilde G$ as a function as reduced pressure  $\tilde P$ for different reduced temperature in Fig.~\ref{fig2}. It can be observed that the red curve, corresponding to $\tilde T < 1$, exhibits swallowtail behavior, with its intersection point signifying a first-order phase transition. However, as $\tilde T$ increases above $1$ (as seen in the blue dotted curve), the curves gradually become smooth, indicating an absence of phase transition in the system, akin to an ideal gas.

\subsubsection{The critical behaviors}
\label{sec3.1.2}
Based on the above discussion, it can be seen that the phase transition of the FRW universe is very similar to that of the van der Waals gas, and therefore, it is necessary to further discuss the critical behavior of the FRW universe. In general, the critical exponents $(\alpha, \varepsilon, \gamma, \delta)$ are defined as \cite{ch9,ch14,ch22,ch23,ch24,ch25}

\vspace{-\baselineskip}
\begin{subequations}
\label{eq15}
\begin{align}
&{C_V}  = T{\left( {\frac{{\partial S}}{{\partial T}}} \right)_V} \propto {\left| \tau  \right|^{ - \alpha }}, \label{eq15-1}\\
&\eta   = \frac{{{V_l} - {V_s}}}{{{V_c}}} \sim \left( {{\omega _l} - {\omega _s}} \right) \propto {\left| \tau  \right|^{\varepsilon }}, \label{eq15-2}\\
&\kappa   =  - \frac{1}{V}{\left( {\frac{{\partial V}}{{\partial P}}} \right)_T} \propto {\left| \tau  \right|^{ - \gamma }}, \label{eq15-3}\\
& P - 1  \propto {\omega ^\delta }, \label{eq15-4}
\end{align}
\end{subequations}
where $\tau \! = \!\tilde T - 1$  and $\omega \! = \!\tilde V - 1$, the labels  ``\emph{s}" and ``\emph{l}" represent ``small" and ``large" states, respectively.

It is easy find that the GUP corrected entropy ${S_{{\text{GUP}}}} = {A \mathord{\left/ {\vphantom {A 4}} \right. \kern-\nulldelimiterspace} 4} + 4\pi \beta \ln \left( {{A \mathord{\left/ {\vphantom {A {{A_0}}}} \right. \kern-\nulldelimiterspace} {{A_0}}}} \right) = {6^{2/3}}{\pi ^{1/3}}{V^{2/3}} + 4\pi \beta \ln \left( {{{{6^{2/3}}{\pi ^{1/3}}{V^{2/3}}} \mathord{\left/ {\vphantom {{{6^{2/3}}{\pi ^{1/3}}{V^{2/3}}} {{A_0}}}} \right. \kern-\nulldelimiterspace} {{A_0}}}} \right)$  is only a function of the thermodynamic volume $V$, hence, one can find that the heat capacity at constant volume ${C_V}$ is zero, which demonstrates first critical exponent as $\alpha=0$.

Next, in order to investigate the other three exponents, one needs to expand the equation of state~(\ref{eq14}) near the critical point, which reads \cite{ch23,ch24,ch25}

\vspace{-\baselineskip}
 \begin{align}
\label{eq16}
\tilde P & = 1 + \frac{8}{{11}}\left( {1 + 2\sqrt 3 } \right)\tau  + \frac{8}{{33}}\left( {3 - 5\sqrt 3 } \right)\tau \omega
\nonumber \\
& + \frac{4}{{297}}\left( {\sqrt 3  - 5} \right){\omega ^3} + \mathcal{O}\left( {t{\omega ^2},{\omega ^4}} \right),
\end{align}
where the term $\mathcal{O}\left( {t{\omega ^2}} \right)$ can be neglected in this expansion is justified by the Eq.~(\ref{eq18}) \cite{ch9} . By employing the Maxwell's equal area law  ${\tilde P^*}\left( {{{\tilde V}_s} - {{\tilde V}_l}} \right) = \int_l^s {\tilde P} d\tilde V$ with the pressure end/starting point of small/large phase  ${\tilde P^*}$, one gets

\vspace{-\baselineskip}
 \begin{align}
\label{eq17}
\frac{8}{{33}}\left( {3 - 5\sqrt 3 } \right)\tau {\omega _l} + \frac{4}{{297}}\left( {\sqrt 3  - 5} \right)\omega _l^3 =
\nonumber \\
\frac{8}{{33}}\left( {3 - 5\sqrt 3 } \right)\tau {\omega _s} + \frac{4}{{297}}\left( {\sqrt 3  - 5} \right)\omega _s^3,
\end{align}
and

\vspace{-\baselineskip}
 \begin{align}
\label{eq18}
\frac{{16}}{{33}}\! \left( {3 - 5\sqrt 3 } \right) \! \tau \! \left( {\omega _l^2 - \omega _s^2} \right) \! \!+\! \frac{{12}}{{297}}\! \left( {\sqrt 3  - 5} \right) \! \!\left( {\omega _l^4 \!- \! \omega _s^4} \right) \! \!=\! 0.
\end{align}
Solving above equations, the nontrivial solutions are ${\omega _s} =  - {3^{5/4}}\sqrt {2\left( {\tilde T - 1} \right)} $  and  ${\omega _l} \! = {3^{5/4}}\sqrt {2\left( {\tilde T - 1} \right)} $, which leads to ${\omega _l} - {\omega _s} \!= 2{\left( { - 18\sqrt 3 \tau } \right)^{1/2}}$, hence, the second critical exponent is  $\varepsilon  = {1 \mathord{\left/ {\vphantom {1 2}} \right. \kern-\nulldelimiterspace} 2}$.

In order to calculate the third critical exponent  $\gamma$, one should differentiates Eq.~(\ref{eq14}), which leads to ${\left. {{{\partial V} \mathord{\left/ {\vphantom {{\partial V} {\partial P}}} \right. \kern-\nulldelimiterspace} {\partial P}}} \right|_T} =  - \left[ {{{33{V_c}} \mathord{\left/ {\vphantom {{33{V_c}} {8\left( {3 - 5\sqrt 3 } \right){P_c}\tau }}} \right. \kern-\nulldelimiterspace} {8\left( {3 - 5\sqrt 3 } \right){P_c}\tau }}} \right] + \mathcal{O}\left( \omega  \right)$. Therefore, one yields

\vspace{-\baselineskip}
\begin{align}
\label{eq20}
\kappa  =  - \frac{1}{V}{\left( {\frac{{\partial V}}{{\partial P}}} \right)_T} =  - \frac{{33}}{{8\left( {3 - 5\sqrt 3 } \right){P_c}\tau}} \propto |\tau{|^{ - \gamma }},
\end{align}
which indicates third exponent is $ \gamma  = 1$. When considering the $\tau=0$, that is  $T = {T_c}$, the fourth critical exponent is

\vspace{-\baselineskip}
\begin{align}
\label{eq21}
\tilde P - 1 = \frac{4}{{297}}\left( {\sqrt 3  - 5} \right){\omega ^3} \Rightarrow \delta  = 3.
\end{align}
Following scaling laws, the four critical exponents $\left(\alpha,\varepsilon,\gamma,\delta \right)$ satisfy the two independent relations

\vspace{-\baselineskip}
\begin{align}
\label{eq21}
&\alpha  + 2\varepsilon  + \gamma  = 2, \alpha  + \varepsilon \left( {1 + \delta } \right) = 2,
\nonumber \\
&\gamma \left( {1 + \delta } \right) = \left( {2 - \alpha } \right)\left( {\delta  - 1} \right),\gamma  = \varepsilon \left( {\delta  - 1} \right),
\end{align}
which coincide with mean field theory. Besidse, according to Eq.~(\ref{eq18}), the volumes of the coexistence small and large phases of FRW universe near the critical point can be expressed as

\begin{figure}[htbp]
\centering
\includegraphics[width=0.32\textwidth]{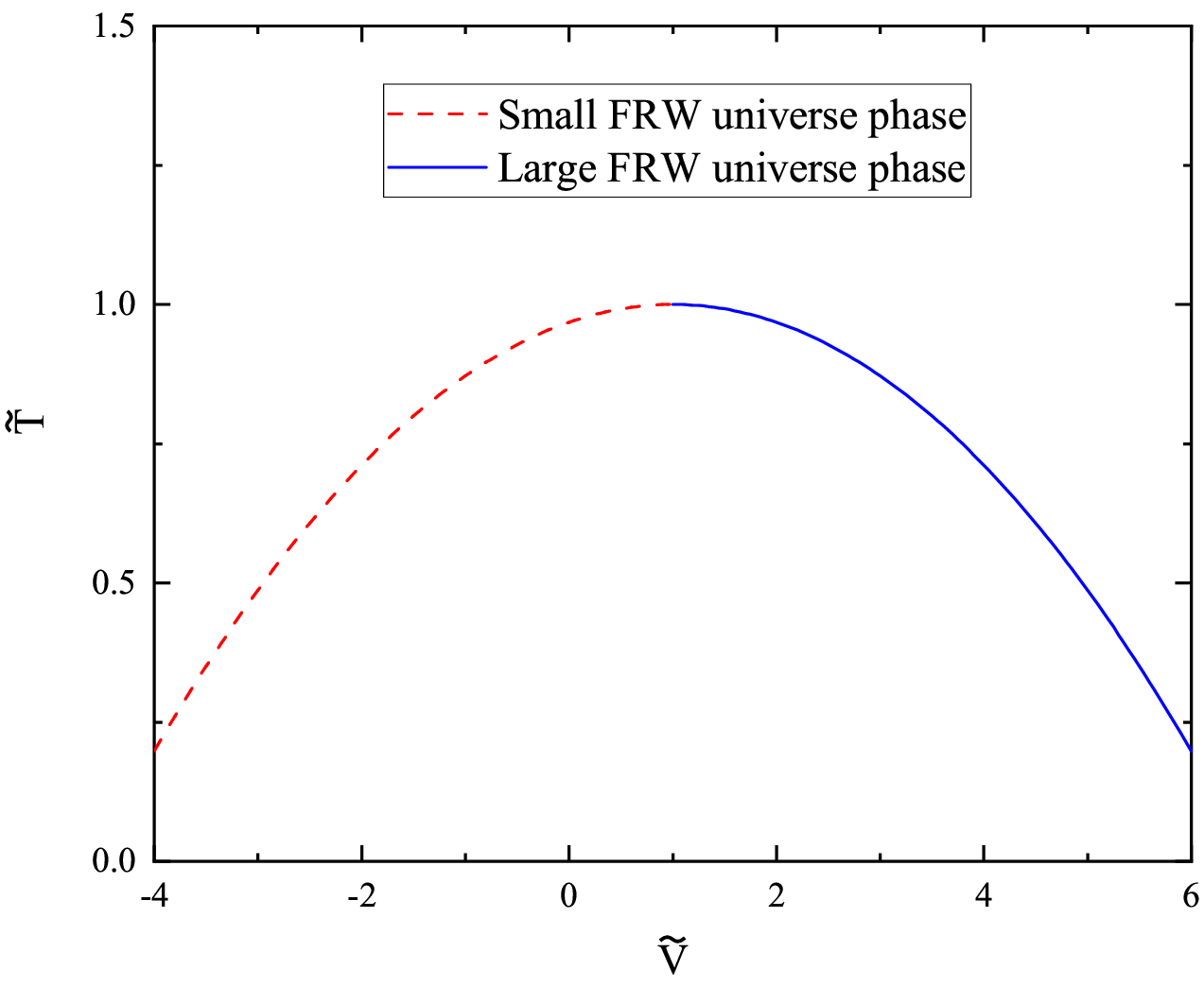}
\caption{\label{fig3} $\tilde T$  as the function as the volumes  $\tilde V$ of the coexistence small/large phases of FRW universe with $\beta>0$.}
\end{figure}

\vspace{-\baselineskip}
\begin{subequations}
\label{eq19}
\begin{align}
&{\tilde V_{\text{s}}} = 1 - {3^{5/4}}\sqrt {2\left( {1 - \tilde T} \right)}, \label{eq19-1}\\
&{\tilde V_l} = 1 + {3^{5/4}}\sqrt {2\left( {1 - \tilde T} \right)}. \label{eq19-2}
\end{align}
\end{subequations}
In Fig.~\ref{fig3}, one can see the phase transition behavior of small/large coexisting phases of FRW universe near the critical point. The red dashed curve representing the large coexisting phase terminates precisely where the blue solid curve representing the small coexisting phase commences, forming a peak at $\left( {\tilde T,\tilde V} \right) = \left( {1,1} \right)$. As a result, the two coexisting states persist in the region  ${\tilde T} < 1$. These observations will be the basis for our analysis of the behavior of thermodynamic curvature.

\subsubsection{The microstructure}
\label{sec3.1.3}
In the last part, we further discuss the microstructure of the FRW universe using phase transitions. To this end, we employ Ruppeiner geometry, a conceptual framework grounded in Riemannian geometry and rooted in the principles of the fluctuation theory of equilibrium thermodynamics. In this language the entropy plays a vital role, and its line element can be expressed in terms of entropy as follows \cite{ch54a}

\vspace{-\baselineskip}
\begin{align}
\label{eq22}
{\text{d}}{l^2} = {g_{\mu \nu }}{\text{d}}{x^\mu }{\text{d}}{x^\nu },
\end{align}
where ${\text{d}}{l^2}$  is the distance between two neighbouring fluctuation states, the fluctuation coordinates $x = \left( {U,V} \right)$ are functions of the internal energy  $U$ and volume  $V$,  ${g_{\mu \nu }} =  - {\partial _{\mu ,\nu }}S$  is the metric element. In the system of FRW universe, the first law of thermodynamics reads

\vspace{-\baselineskip}
\begin{align}
\label{eq23}
{\text{d}}S = \frac{1}{T}{\text{d}}U + \frac{P}{T}{\text{d}}V.
\end{align}
By comparing Eq.~(\ref{eq22}) with Eq.~(\ref{eq23}), and the conjugate quantities corresponding to $x$  are ${y_\mu } \!=\! \partial S/\partial {x^\mu } = \left( {{1 \mathord{\left/ {\vphantom {1 {T,{{ - V} \mathord{\left/ {\vphantom {{ - V} T}} \right. \kern-\nulldelimiterspace} T}}}} \right. \kern-\nulldelimiterspace} {T,{{ - V} \mathord{\left/ {\vphantom {{ - V} T}} \right. \kern-\nulldelimiterspace} T}}}} \right)$. Therefore, one has the following relations

\vspace{-\baselineskip}
\begin{subequations}
\label{eq24}
\begin{align}
&{\text{d}}\left( {\frac{1}{T}} \right) = {\left( {\frac{{{\partial ^2}S}}{{\partial {U^2}}}} \right)_V}{\text{d}}U + \left( {\frac{{{\partial ^2}S}}{{\partial U\partial V}}} \right){\text{d}}V, \label{eq24-1}\\
&{\text{d}}\left( {\frac{P}{T}} \right) = {\left( {\frac{{{\partial ^2}S}}{{\partial {V^2}}}} \right)_U}{\text{d}}V + \left( {\frac{{{\partial ^2}S}}{{\partial U\partial V}}} \right){\text{d}}U. \label{eq24-2}
\end{align}
\end{subequations}
Now, the line element~(\ref{eq22}) can be rewritten as

\vspace{-\baselineskip}
\begin{align}
\label{eq25}
{\text{d}}{l^2}{\text{ }} & =  - {\text{d}}\left( {\frac{1}{T}} \right){\text{d}}U - {\text{d}}\left( {\frac{P}{T}} \right){\text{d}}V
\nonumber \\
& = \frac{1}{{{T^2}}}{\text{d}}T{\text{d}}U - \frac{1}{T}{\text{d}}P{\text{d}}V + \frac{P}{{{T^2}}}{\text{d}}T{\text{d}}V.
\end{align}
Based on the relation  ${\text{d}}U \!=\! {C_V}{\text{d}}T + \left[ {T{{\left( {{{\partial P} \mathord{\left/ {\vphantom {{\partial P} {\partial T}}} \right.
 \kern-\nulldelimiterspace} {\partial T}}} \right)}_V} - P} \right]{\text{d}}V$ and  ${\text{d}}P \!=\! {\left( {{{\partial P} \mathord{\left/ {\vphantom {{\partial P} {\partial T}}} \right. \kern-\nulldelimiterspace} {\partial T}}} \right)_V}{\text{d}}T + {\left( {{{\partial P} \mathord{\left/ {\vphantom {{\partial P} {\partial T}}} \right.
 \kern-\nulldelimiterspace} {\partial T}}} \right)_T}{\text{d}}V$, the line element can be further expressed in term  $\left( {T,V} \right)$ as \cite{ch15}

 \vspace{-\baselineskip}
\begin{align}
\label{eq26}
{\text{d}}{l^2} = \frac{{{C_V}}}{{{T^2}}}{\text{d}}{T^2} - \frac{{{{\left( {{\partial _V}P} \right)}_T}}}{T}{\text{d}}{V^2}.
\end{align}
In order to study the thermodynamic geometry as well as the microstructure of the FRW universe, it is necessary to analyze its scalar curvature  $R$. However, it is easy proved that the entropy ${C_V}$  depends only on the volume, so that the thermodynamic line element~(\ref{eq26}) is singular, which means that the classical definition of  $R$ is always divergent. To avoid this predicament, a new reduced curvature scalar that so called thermodynamic curvature scalar is defined as follows \cite{ch14b+,ch15b+,ch15}

\vspace{-\baselineskip}
\begin{align}
\label{eq27}
&{R_N} = R{C_V}
\nonumber \\
&= \frac{{\left( {{\partial _V}P} \right)_T^2 - {T^2}{{\left( {{\partial _{T,V}}P} \right)}^2} + 2{T^2}{{\left( {{\partial _V}P} \right)}_T}\left( {{\partial _{T,T,V}}P} \right)}}{{2\left( {{\partial _V}P} \right)_T^2}}.
\end{align}
By taking this probe, one can easily analyze the thermodynamic geometry as well as the microstructure of the FRW universe. According to Eq.~(\ref{eq14}) and Eq.~(\ref{eq27}), the reduced thermodynamic curvature scalar reads

\begin{figure}[htbp]
\centering
\subfigure[]{
\begin{minipage}{0.32\textwidth}
\includegraphics[width=\textwidth]{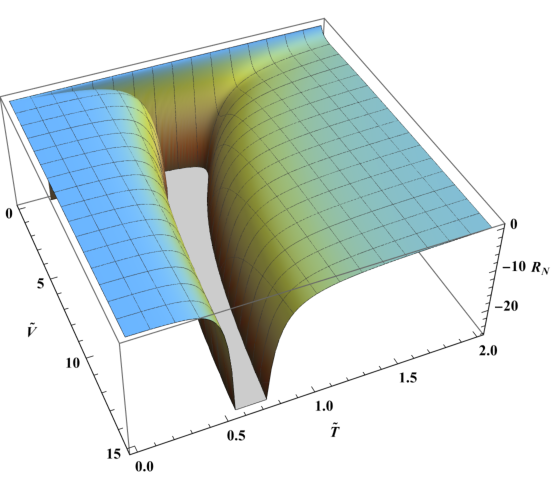}
\label{fig4-a}
\end{minipage}
}
\subfigure[]{
\begin{minipage}{0.32\textwidth}
\includegraphics[width=\textwidth]{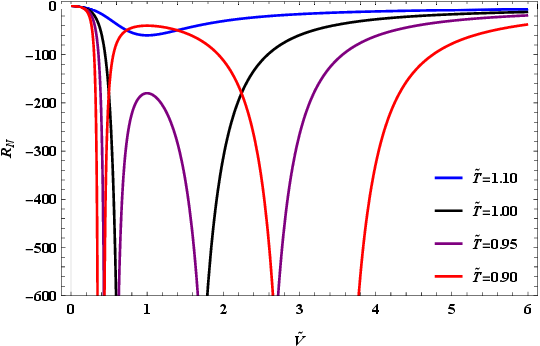}
\label{fig4-b}
\end{minipage}
}
\subfigure[]{
\begin{minipage}{0.32\textwidth}
\includegraphics[width=\textwidth]{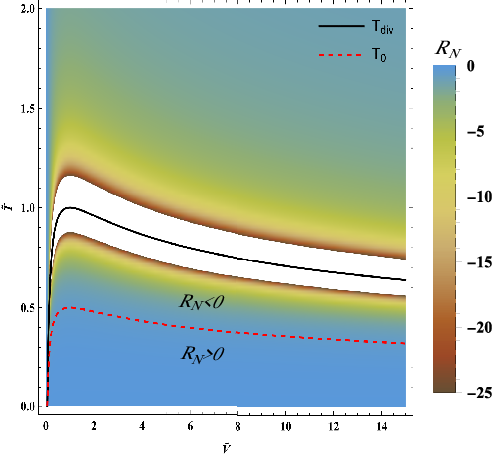}
\label{fig4-c}
\end{minipage}
}
\caption{The reduced scalar curvature versus the reduce volume and reduce temperature with $\beta>0$.}
\label{fig4}
\end{figure}

\vspace{-\baselineskip}
\begin{align}
\label{eq28}
{R_N} = \frac{{\mathcal{A} - 2\tilde T \mathcal{B}}}{{2{\mathcal{C}^2}}},
\end{align}
where $\mathcal{A} \! = \! 97  \!+ \! 56\sqrt 3 \! - \!  2( {627 \! + \! 362\sqrt 3 } ){{\tilde V}^{2/3}} \! + \! 3( {1351  \!+ \! 780\sqrt 3 }){{\tilde V}^{4/3}}$, $\mathcal{B} \! = \! ({724 + 418\sqrt 3 }) \tilde V \! + \! ({2340 \! + \! 1351\sqrt 3 }){{\tilde V}^{5/3}} \! - \! ({168 + 97\sqrt 3 }){{\tilde V}^{1/3}}$, and $\mathcal{C} \! = \! 7 \! + \! 4\sqrt 3  \! + \! ({12 \! + \! 7\sqrt 3 })\tilde T{{\tilde V}^{1/3}}\!  - \! ({45 \! +\!  26\sqrt 3 }){{\tilde V}^{2/3}} \! +\! ({26 \! + \! 15\sqrt 3 })\tilde T\tilde V$. Now, the reduced thermodynamic curvature scalar only related to the $\tilde V$  and  $\tilde T$, rather than the GUP parameter, property is similar to reduced equation of state. Moreover, we depict the behavior of $R_N$ in Fig.~\ref{fig4}. One can observe that  $R_N$ is near zero for most of the parameter space (see Fig.~\ref{fig4-a}). However, near the temperature

\vspace{-\baselineskip}
\begin{align}
\label{eq29}
{\tilde T_{{\text{div}}}} = \frac{{3\left( {2 + \sqrt 3 } \right){V^{2/3}} - \sqrt 3 }}{{3{V^{1/3}} + \left( {3 + 2\sqrt 3 } \right)V}},
\end{align}
the curve of  $R_N$ decreases dramatically and eventually reaches negative infinity, indicating that the microstructure of the universe is changing rapidly near  ${\tilde T_{{\text{div}}}} $. As depicted in Fig.~\ref{fig4-b},  one can see the divergence behavior of ${R_N} $. For $\tilde T > 1$, no divergence is observed. For $\tilde T = 1$,  a single divergence point is present. However, when $\tilde T$ falls below $1$, this divergence point bifurcates, with the two points moving towards the high and low volume regions, respectively. By solving $R_N=0$ allows us to deduce the sign-changing curve

\vspace{-\baselineskip}
\begin{align}
\label{eq29+}
{T_0} = \frac{{\left( {26 + 15\sqrt 3 } \right)\left( {3 - 2\sqrt 3  + 3{{\tilde V}^{2/3}}} \right)}}{{6\left( {7 + 4\sqrt 3 } \right){{\tilde V}^{1/3}} + \left( {90 + 52\sqrt 3 } \right)\tilde V}},
\end{align}
which implies a transition between attractive and repulsive interactions in the microstructure. In Fig.~\ref{fig4-c}, we illustrate  the sign-changing (red dashed) curve and the so-called spinodal (black solid) curve corresponds to the divergent temperature ${\tilde T_{{\text{div}}}}$. Above the sign-changing curve, the ${R_N}$ is negative, which implies that attractive interactions dominate in this region. However, below the sign-changing curve the ${R_N}$ becomes positive, indicating the repulsive interactions dominate. Therefore, the microstructure of  FRW universe exhibits repulsive interactions at low temperature and attractive interactions at high temperature. Besides, one can see that the black solid curve is far from the red dashed curve in cases where the volume is not very small, which suggests that the change in ${R_N}$ in region below the red dashed curve is very close to zero, and consequently, the attractive interaction there is weak.

As we know, the reduced thermodynamic curvature scalar along the small-phase and large-phase coexistence curves near the critical points would provide some universal properties. To examine whether the FRW universe in the framework of GUP has these properties, we insert the volumes of the coexistence small and large phases of the FRW universe~(\ref{eq19}) into Eq.~(\ref{eq29}),  the thermodynamic curvature scalar around the critical point is given by

\vspace{-\baselineskip}
\begin{subequations}
\label{eq30}
\begin{align}
R_N^{{s}} =  - \frac{1}{{8{\tau ^2}}} + \frac{{\sqrt {27 + 42\sqrt 3 } }}{{4{{\left( { - \tau } \right)}^{3/2}}}} + \mathcal{O}\left( {{\tau ^{ - 1}}} \right), \label{eq30-1}\\
R_N^{{l}} =  - \frac{1}{{8{\tau ^2}}} - \frac{{\sqrt {27 + 42\sqrt 3 } }}{{4{{\left( { - \tau } \right)}^{3/2}}}} + \mathcal{O}\left( {{\tau ^{ - 1}}} \right). \label{eq30-2}
\end{align}
\end{subequations}
Thus, the thermodynamic curvature scalar has a critical exponent $2$. When, by ignoring the high orders terms of Eq.~(\ref{eq30-1}) and Eq.~(\ref{eq30-2}), an interesting limiting expression can be obtained that

\vspace{-\baselineskip}
\begin{align}
\label{eq31}
\mathop {\lim }\limits_{\tau  \to 0} {R_N}{\tau ^2} =  - \frac{1}{8},
\end{align}
which indicates that the divergence behavior of thermodynamic curvature scalar of FRW universe in the GUP framework is characterized by a dimensionless constant  $ - {1 \mathord{\left/ {\vphantom {1 8}} \right. \kern-\nulldelimiterspace} 8}$. This result is consistent with the previous works, such as AdS black holes in extended phase space and VdW system \cite{ch15,ch17}.

\subsection{The phase transition, critical behaviors and microstructure for $\beta  < 0$  scenario}
\label{sec3.2}
\subsubsection{The phase transition}
\label{sec3.2.1}
In this subsection, we examine the properties of the phase transition, criticality and microstructure of the FRW universe with the negative GUP parameter. By using Eq.~(\ref{eq8})-Eq.~(\ref{eq10}), the critical radius of apparent horizon, temperature, and pressure for $\beta<0$ are

\vspace{-\baselineskip}
\begin{align}
\label{eq32}
{R_c} & = 2\sqrt {\left( {3 - 2\sqrt 3 } \right)\beta }, {V_c}   = \frac{4}{3}\pi R_c^3,
\nonumber \\
{T_c} & = \frac{1}{{4\pi \sqrt {\left( {9 - 6\sqrt 3 } \right)\beta } }}, {P_c} =  - \frac{{15 + 8\sqrt 3 }}{{576 \pi \beta }},
\end{align}
and the three critical points give the ratio

\vspace{-\baselineskip}
\begin{align}
\label{eq33}
\chi=\frac{{{\nu _c}{P_c}}}{{{T_c}}} = \frac{{6 + \sqrt 3 }}{{24}} \approx 0.64.
\end{align}
Unlike the Eq.~(\ref{eq12}), the critical ratio is much larger than ${3 \mathord{\left/ {\vphantom {3 8}} \right. \kern-\nulldelimiterspace} 8}$, which indicates that the phase transition, criticality and microstructure of the FRW universe with the negative GUP parameter may different from those of VdW system and the charged AdS black hole in extended phase space. To check this conjecture, it is necessary to derive thermodynamic equation of state. By using Eq.~(\ref{eq12}), Eq.~(\ref{eq13}) and Eq.~(\ref{eq32}), the modified reduced thermodynamic equation of state is provided as follows:

\vspace{-\baselineskip}
\begin{align}
\label{eq34}
\tilde P \! = \! \frac{{6\left( {3 - 2\sqrt 3 } \right){{\tilde V}^{2/3}} + 4\tilde T\left[ {\sqrt 3 {{\tilde V}^{1/3}} + 3\left( { - 2 + \sqrt 3 } \right)\tilde V} \right] - 3}}{{\left( {4\sqrt 3  - 9} \right){{\tilde V}^{4/3}}}}.
\end{align}

\begin{figure}[htbp]
\centering
\includegraphics[width=0.32\textwidth]{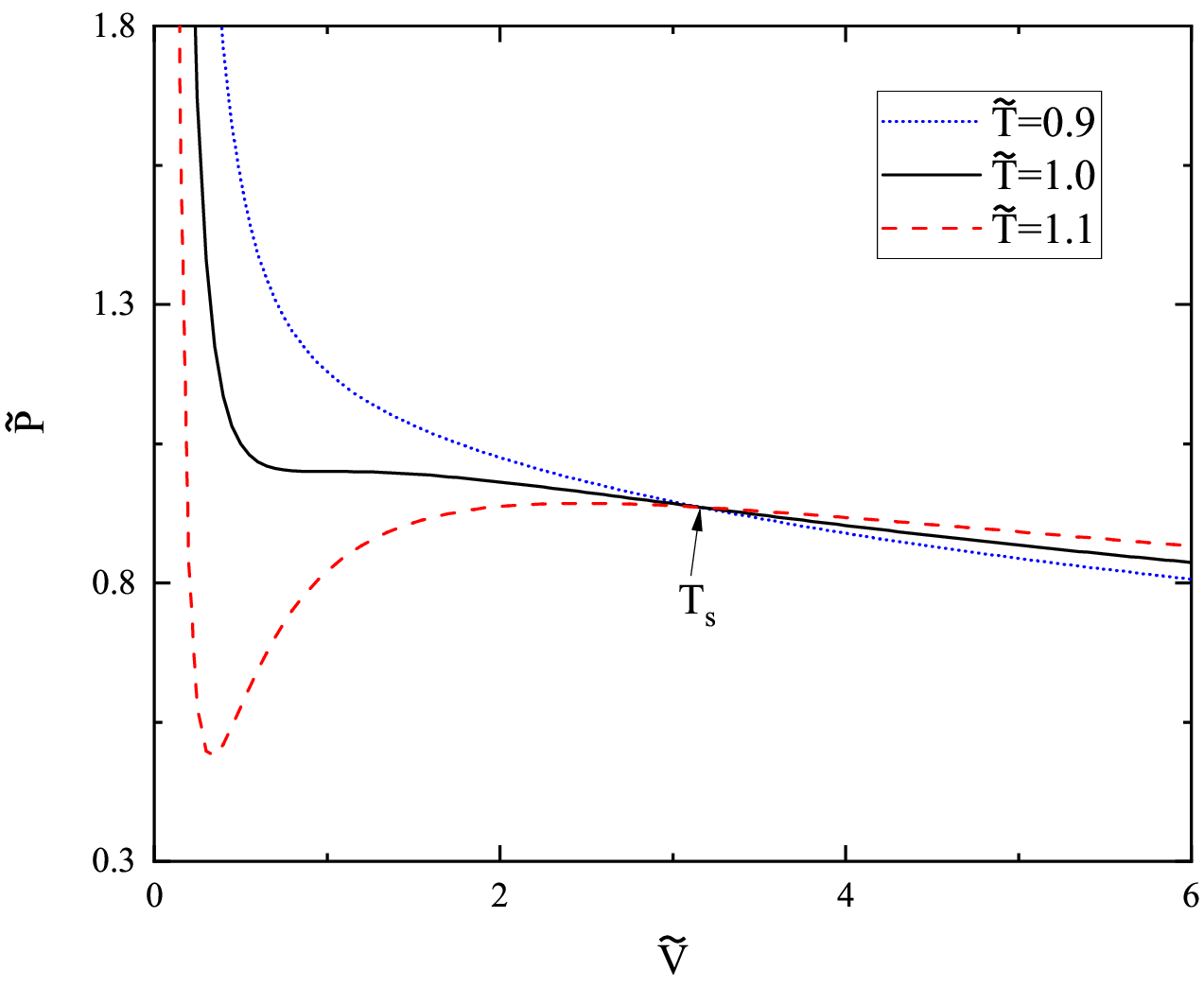}
\caption{\label{fig6}  Variation in reduced pressure with reduced volume for different reduced temperature with $\beta<0$.}
\end{figure}

According to the equation above, the  relationship between $\tilde P$  and $\tilde V$  for different  $\tilde T$  is depicted in Fig.~\ref{fig6}.

In Fig.~\ref{fig6}, the behavior of red dashed curve shows that the phase transition occurs for $\tilde T > 1$, while for $\tilde T < 1$, the blue isotherm resembles an ideal gas. Those behaviors are in contrast to the trends in Fig.~\ref{fig1}. Besides, one can see that all the curves intersect at point $T_s$, where is characterized by a pressure independent of temperature, and interpreted as a ``thermodynamic singularity'' \cite{ch24,ch25,ch59,ch60,ch61}. All the difference indicates that the GUP with a negative parameter leads to a phase transition behavior that is different from that of VdW system and most black hole systems in extended phase space.

Moreover, the observed phase transition phenomena of the FRW universe in the $P-V$ plane, as presented above, can also be expressed through the reduced "Gibbs free energy-pressure" relation, as shown in Eq.~(\ref{eq15}).
\begin{figure}[htbp]
\centering
\includegraphics[width=0.32\textwidth]{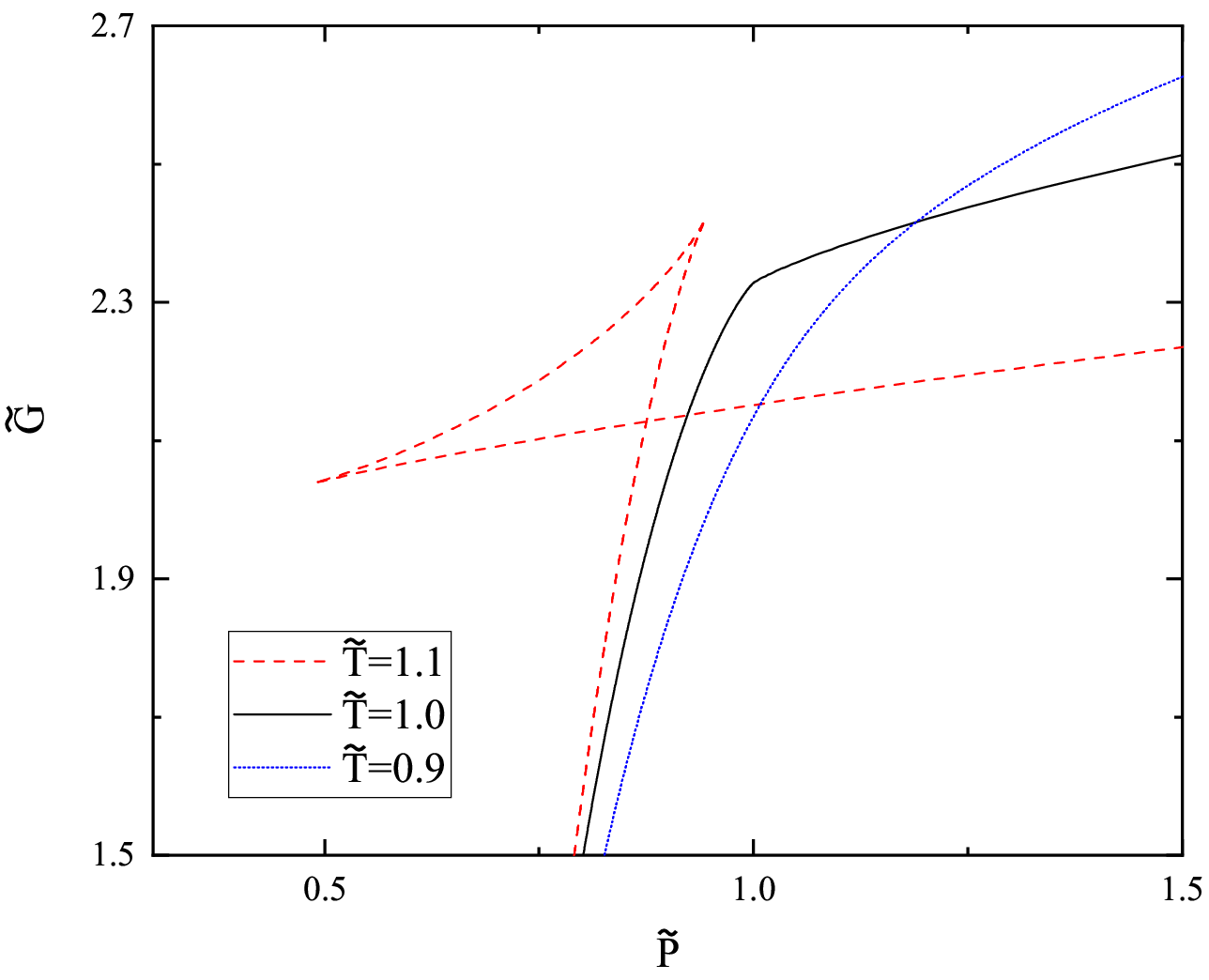}
\caption{\label{fig7}  Variation in reduced Gibbs free energy with reduced pressure for different reduced temperature with $\beta<0$.}
\end{figure}
From Fig.~\ref{fig7}, it can be observed that the smooth blue dotted curve for $\tilde T < 1$ corresponds to the behavior of an ideal gas in the $\tilde G - \tilde P$ plane, indicating that there is no phase transition in the system. When $\tilde T = 1$, the black solid curve is no longer smooth and has a breaking point, meaning that the system is in a critical state. For $\tilde T > 1$, the reduced Gibbs free energy exhibits a swallow-tail behavior (red dashed curve) in the $\tilde G - \tilde P$ plane, which indicates that there is a two-phase coexistence state. Therefore, the intersection represents a first-order phase transition.

\subsubsection{The critical behaviors}
\label{sec3.2.2}
Next, by defining $\tau  = \tilde T - 1$  and  $\omega  = \tilde V - 1$, the equation of state~(\ref{eq34}) near the critical point can be rewritten as follows

\vspace{-\baselineskip}
\begin{align}
\label{eq35}
\tilde P & = 1 + \frac{8}{{11}}\left( {1 - 2\sqrt 3 } \right)\tau  + \frac{8}{{33}}\left( {3 + 5\sqrt 3 } \right)\tau \omega
\nonumber \\
& - \frac{4}{{297}}\left( {5 + \sqrt 3 } \right){\omega ^3} + \mathcal{O}\left( {\tau{\omega ^2},{\omega ^4}} \right),
\end{align}
where the truncation of the series is justified by formula Eq.~(\ref{eq36-2}) below \cite{ch9}. Based on Eq.~(\ref{eq15}) and Eq.~(\ref{eq35}), and  then performing simple calculations, the critical exponents are obtained as

\vspace{-\baselineskip}
\begin{subequations}
\label{eq36}
\begin{align}
{C_V} = 0 &\Rightarrow \alpha  = 0, \label{eq36-1}\\
\eta  = 2{\left( { - 18\sqrt 3 \tau } \right)^{1/2}} &\Rightarrow \varepsilon  = \frac{1}{2}, \label{eq36-2}\\
\kappa  =  - \frac{{33}}{{8\left( {3 - 5\sqrt 3 } \right){P_c}t}} &\Rightarrow \gamma  = 1, \label{eq36-3}\\
\tilde P - 1 = \frac{4}{{297}}\left( {\sqrt 3  - 5} \right){\omega ^3} &\Rightarrow \delta  = 3, \label{eq36-4}
\end{align}
\end{subequations}
which are consistent with the predications from the mean field theory, hence, they are also satisfy scaling laws~(\ref{eq23}). Besides, by using the Maxwell's equal area law, the volumes of the coexistence small and large phases of FRW universe around the critical point can be expressed as

\begin{figure}[htbp]
\centering
\includegraphics[width=0.32\textwidth]{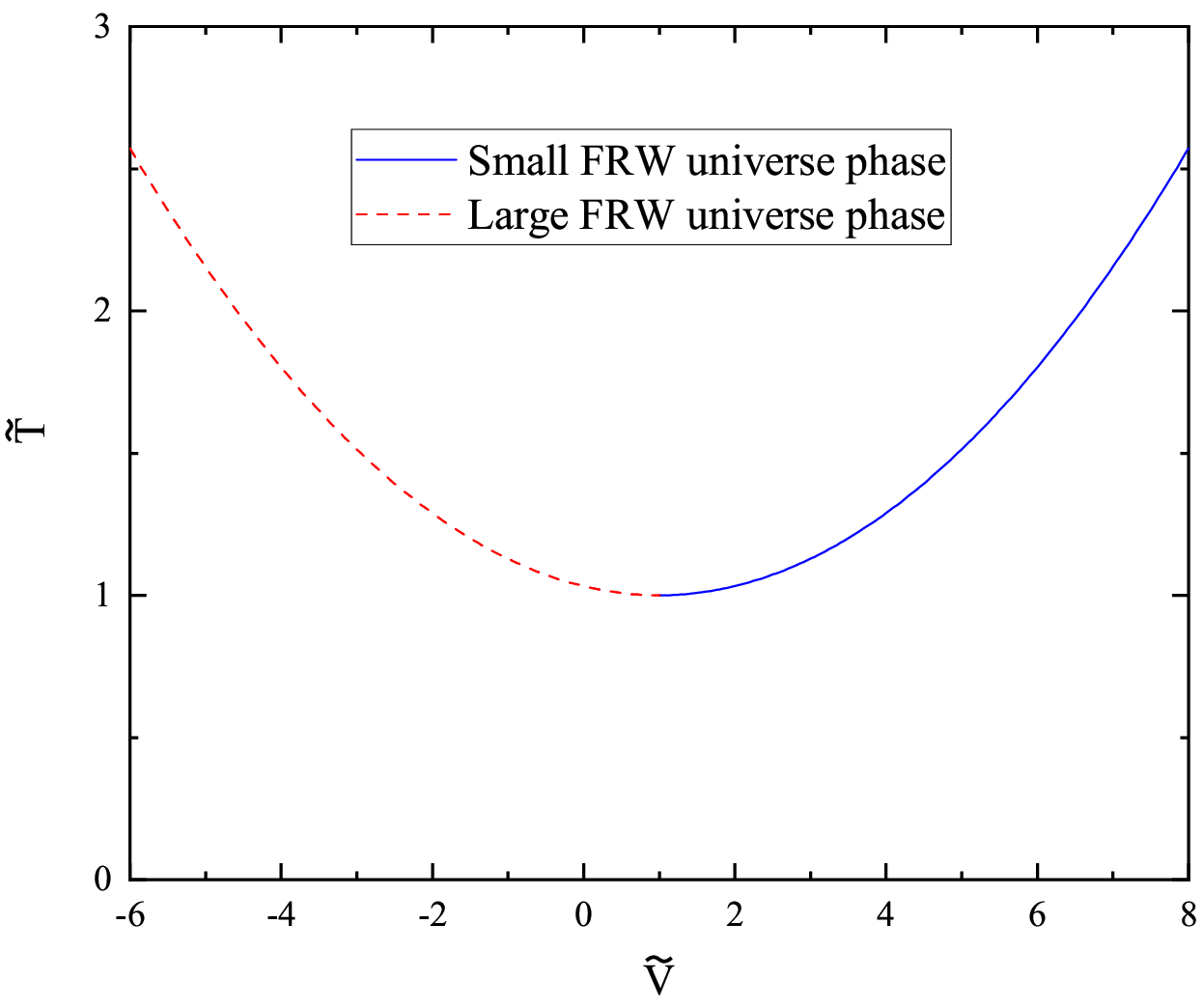}
\caption{\label{fig8}  $\tilde T$  as the function as the volumes  $\tilde V$ of the coexistence small/large phases of FRW universe with $\beta<0$.}
\end{figure}

\vspace{-\baselineskip}
\begin{subequations}
\label{eq37}
\begin{align}
{\tilde V_s} = 1 - {3^{5/4}}\sqrt {2\left( {\tilde T - 1} \right)} ,  \label{eq37-1}\\
{\tilde V_l} = 1 + {3^{5/4}}\sqrt {2\left( {\tilde T - 1} \right)}.  \label{eq37-2}
\end{align}
\end{subequations}
According to Eq.~(\ref{eq37}), we show the temperature  ${\tilde T}$ as a function of volumes ${\tilde V}$  in Fig.~\ref{fig8}. The red and blue curves represent the large coexisting phase and small coexisting phase, respectively. The two curves intersect at the point $\left( {\tilde T,\tilde V} \right) = \left( {1,1} \right)$  where the volume is minimized, this the opposite of that is shown in Fig.~\ref{fig2}. Therefore, the two coexisting states would exist in the region $\tilde T > 1$.

\subsubsection{The microstructure}
\label{sec3.2.3}
Based on Eq.~(\ref{eq22})-Eq.~(\ref{eq25}), one can express the line element of the Ruppeiner geometry as ${\text{d}}{l^2} = {{{C_V}} \mathord{\left/ {\vphantom {{{C_V}} {{T^2}}}} \right. \kern-\nulldelimiterspace} {{T^2}}}{\text{d}}{T^2} - {{{{\left( {{\partial _V}P} \right)}_T}} \mathord{\left/ {\vphantom {{{{\left( {{\partial _V}P} \right)}_T}} T}} \right. \kern-\nulldelimiterspace} T}{\text{d}}{V^2}$ \cite{ch15,ch54a}, and the corresponding the expression of scalar curvature ${R_N}$  is the same as Eq.~(\ref{eq27}). Thus, by putting Eq.~(\ref{eq35}) into Eq.~(\ref{eq27}), the scalar curvature has been obtained \cite{ch14b+,ch15b+,ch15}

\vspace{-\baselineskip}
\begin{align}
\label{eq38}
{R_N} = \frac{{\mathcal{C} + 2\tilde T \mathcal{D}}}{{2{\mathcal{E}^2}}},
\end{align}
where $\mathcal{C}={1 + \left( {4\sqrt 3  - 6} \right){{\tilde V}^{2/3}} + 3\left( {7 - 4\sqrt 3 } \right){{\tilde V}^{4/3}}}$, $\mathcal{D}={2\left( {\sqrt 3  - 2} \right)\tilde V + \left( {7\sqrt 3  - 12} \right){{\tilde V}^{5/3}} - \sqrt 3 {{\tilde V}^{1/3}}}$, and $\mathcal{E}=\sqrt 3 \tilde T{{\tilde V}^{1/3}} $ $+ \left( {3 - 2\sqrt 3 } \right){{\tilde V}^{2/3}} + \left( {\sqrt 3  - 2} \right)\tilde T\tilde V - 1$. The result above is only dependent on ${\tilde V}$  and  ${\tilde T}$, which is consistent with the conclusion of Eq.~(\ref{eq28}). However, the two expressions are not the same, which indicate that the positive/negative GUP parameters lead to two different thermodynamic systems.

\begin{figure}[htbp]
\centering
\subfigure[]{
\begin{minipage}{0.32\textwidth}
\includegraphics[width=\textwidth]{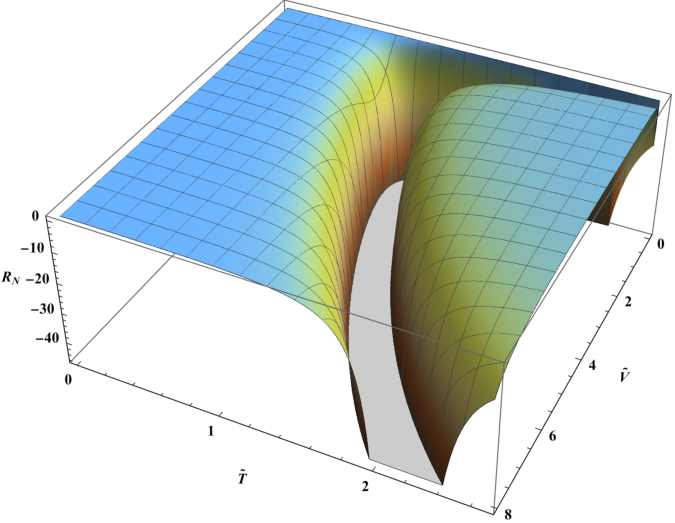}
\label{fig9-a}
\end{minipage}
}
\subfigure[]{
\begin{minipage}{0.32\textwidth}
\includegraphics[width=\textwidth]{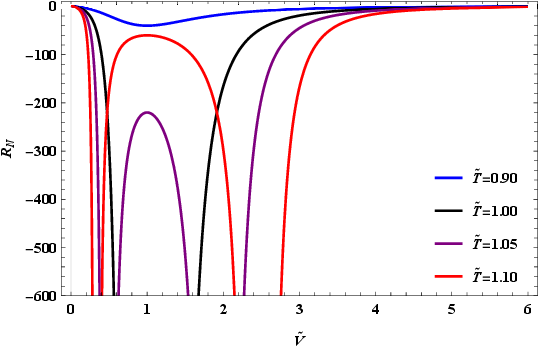}
\label{fig9-b}
\end{minipage}
}
\subfigure[]{
\begin{minipage}{0.32\textwidth}
\includegraphics[width=\textwidth]{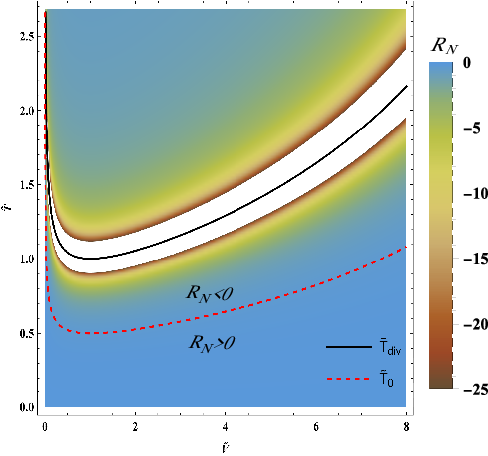}
\label{fig9-c}
\end{minipage}
}
\caption{The behavior of  $R_N$ along the small and large coexistence phases for the FRW universe with $\beta<0$.}
\label{fig9}
\end{figure}

In Fig.~\ref{fig9}, we plot the thermodynamic curvature scalar  ${R_N}$ as a function of $\tilde T$  and $\tilde V$. Fig.~\ref{fig9-a} shows that the thermodynamic curvature scalar is around $0$ for most $(T, V)$ intervals. However, it diverges at temperature

\vspace{-\baselineskip}
\begin{align}
\label{eq39}
{\tilde T_{{\text{div}}}} = \frac{{1 + \left( {2\sqrt 3  - 3} \right){{\tilde V}^{2/3}}}}{{\sqrt 3 {{\tilde V}^{1/3}} + \left( {\sqrt 3  - 2} \right)\tilde V}},
\end{align}
which means that the interaction of microstructure of FRW universe changes rapidly in the vicinity of  ${\tilde T_{{\text{div}}}}$. Moreover, from Fig.~\ref{fig9-b}, it can be seen that there is no divergent point for low temperature $\tilde T < 1$. However, the divergent points appear when the reduced temperature is above $1$. For $\tilde T = 1$, one can see only one negative divergent point at $\tilde V = 1$. For $\tilde T > 1$, there are two negative divergent points, which are moved toward the high volume region and the low volume region, respectively, as temperature increases. In Fig.~\ref{fig9-c}, the red dashed curve represents sign-changing curve, which is expressed as

\vspace{-\baselineskip}
\begin{align}
\label{eq40}
{T_0} \! = \! \frac{{\left( {6 - 4\sqrt 3 } \right){V^{2/3}} \! +  \! 3\left( { - 7 + 4\sqrt 3 } \right){V^{4/3}} \! -  \!1}}{{4\left( {\sqrt 3 \!  -  \!2} \right)V \! + \! 2\left( {7\sqrt 3  - 12} \right){V^{5/3}} \! -  \!2\sqrt 3 {V^{1/3}}}}.
\end{align}
Below the red dashed curve, $R_N$ is larger than $0$, meaning that the microstructure of this region is dominated by repulsive interaction. For the region above the red dashed curve, one has $R_N > 0$, indicating that the microstructure in this region is dominated by attractive interaction. However, the red dashed curve is far from the black curve. Therefore, the $R_N$ below the red dashed curve is very close to $0$, indicating that the repulsive interaction of the system is very weak.

Finally, by inserting Eq.~(\ref{eq37}) into Eq.~(\ref{eq38}), the divergence behavior of thermodynamic scalar curvature along the coexistence small and large
phases can be seen from the following expressions

\vspace{-\baselineskip}
\begin{subequations}
\label{eq40}
\begin{align}
R_N^s =  - \frac{1}{{8{\tau ^2}}} + \frac{{\sqrt {3\left( {14\sqrt 3  - 9} \right)} }}{{4{\tau ^{3/2}}}} + \mathcal{O}\left( {{\tau ^{ - 1}}} \right) ,  \label{eq40-1}\\
R_N^l =  - \frac{1}{{8{\tau ^2}}} - \frac{{\sqrt {3\left( {14\sqrt 3  - 9} \right)} }}{{4{\tau ^{3/2}}}} + \mathcal{O}\left( {{\tau ^{ - 1}}} \right). \label{eq40-2}
\end{align}
\end{subequations}
Obviously, the above result explicitly reveal that $R_N$  has a critical exponent 2, and the divergence behavior is characterized by a dimensionless constant $ - {1 \mathord{\left/ {\vphantom {1 8}} \right. \kern-\nulldelimiterspace} 8}$  for  $\mathop {\lim }\limits_{\tau  \to 0} {R_N}{\tau ^2}$. This results agree with what we obtained in Eq.~(\ref{eq30}) and Eq.~(\ref{eq31}), showing that the properties are universal.

\section{Conclusion}
\label{sec4}
In this paper,  we investigated the phase transition, critical behavior and microstructure of the FRW universe in the framework of a new higher order GUP~(\ref{eq1}).
By defining the work density as a thermodynamic pressure ($W:=P$), the thermodynamic equation of state for the FRW universe is obtained. Then, with simple calculations, we find that there are two sets of possible critical quantities (Eq.~(\ref{eq11}) and Eq.~(\ref{eq32})) in the system, and they hold for GUP parameter $\beta$ larger than zero and less than zero, respectively.  Considering the flexibility of the GUP to accommodate both $\beta>0$ and $\beta<0$, we thereby discuss the thermodynamic properties in these dual scenarios.

For the $\beta>0$ case, we first calculated the ratio of critical quantities $\chi$. As elucidated in Eq.~(\ref{eq12}), it is found that the ratio is very close to those of VdW system and charged AdS black holes in extended phase space, which indicates the thermodynamic properties of FRW universe with positive GUP parameter like those in the previous studies~\cite{ch8+,ch9,ch10,ch11,ch12,ch13,ch14,ch15,ch16,ch17}.  In order to confirm this conjecture, we then discuss the relationship of ``Pressure-Volume'' and  `Free Gibbs energy-Temperature'' near the reduced critical temperature. The results showed that the FRW universe  occurs  a first-order $P-V$ phase transition when the temperature less than the critical temperature, that is $T < {T_c}$ or $\tilde T<1$, and it becomes to the ideal gas when $T > {T_c}$. Subsequently, we derived the critical exponents $\left(\alpha,\varepsilon,\gamma,\delta \right)$ around the critical points, outcomes revealed that the critical exponents in the case $\beta>0$  remain consistent with the predications from the mean field theory. Finally, utilizing the Ruppeiner geometry, we derived the thermodynamic curvature scalar $R_N$, exploring its sign-changing curve and spinodal curve. It was observed that the sign-changing curve and spinodal curve initially increase with growing volume, reaching a maximum value before decreasing with further volume increase. The $R_N$ is negative above the sign-changing curve, which implies the attractive interaction dominate there. For the region below the red dashed curve, ${R_N}$ becomes positive. Conversely, in the region below the red dashed curve, ${R_N}$ becomes positive. However, given the considerable distance from the spinodal curve, the value of ${R_N}$ in this region approaches zero, indicating the presence of weak repulsive interactions.   Conversely, in the region below the red dashed curve, ${R_N}$ becomes positive. However, given the considerable distance from the spinodal curve, the value of ${R_N}$ in this region approaches zero, indicating the presence of weak repulsive interactions.  Based on the above results, it is suggested that the thermodynamic properties and phase transition behavior of FRW universe with positive GUP parameter are very similar to those of the VdW systems and charged AdS black holes in the extended phase space. Considering that the higher-order GUP~(\ref{eq1}) is more complex than the KMM and ADV models, it is believed that there is also a phase transition in the FRW universe in the framework of the KMM model or ADV model, which behaves in a similar way to the  VdW systems.

Conversely, when considering the case $\beta<0$, some different results are yielded. First, Eq.~(\ref{eq33}) showed that the ratio of critical quantities for $\beta<0$ is much larger (almost twice as large) than that of VdW system and charged AdS black holes in extended phase space. Then, from Fig.~\ref{fig6} and  Fig.~\ref{fig7}, one can see that the $P-V$ phase transitions and corresponding swallow-tail behavior of the FRW universe takes place at $T > {T_c}$, while the ideal gas behavior surfaces at $T < {T_c}$, which is the opposite to that of $\beta>0$ case. What adds a layer of interestingly is the emergence of a ``thermodynamic singularity'' in this case, a feature absent in VdW system and most of black holes system. Besides, the analysis of the coexistence curve shows that it has only one singular minimum, signifying that the size phase of the universe exists solely above this minimum, which is also the opposite to that of $\beta>0$ case.  For the  microstructures of the FRW system, it is found  an increase in volume leads to a decreasing and then increasing behavior for both the sign-changing curve and the spinodal curve. Above the sign-changing curve, $R_N>0$, whereas below the sign-changing curve, $R_N<0$. Consequently, the microstructure of the FRW universe is characterized by attractive interactions in the low-temperature region and repulsive interactions in the high-temperature region. These results illustrated that for FRW univers with negative GUP parameter, its thermodynamic properties and phase transition behavior different from those of the VdW systems and most AdS black hole systems, which implies that the interaction between GUP with negative parameters and gravity is different from the case with positive parameters. Nevertheless, it does not mean that our study is problematic,   since they resemble those derived from effective scalar field theory in Refs.~\cite{ch22,ch23,ch24,ch25}. This indicates that a possible deeper connection between effective scalar field theory and GUP with negative parameter, which will be an issue for our future research.

In summary, our findings showed that when considering GUP gravitational effects, there is a phase transition behavior in the FRW universe, the exact properties of which depend on the interaction of GUP with spacetime and gravity. Furthermore, considering that these results are obtained under the same GUP~(\ref{eq1}), they should be considered as a whole. Therefore, our work can show more information compared to the thermodynamic phase transition of black holes or those of the universe under modified gravity. Besides, considering the numerous existing GUP models, we anticipate that these models could also induce phase transitions in the universe. Moreover, it is believed that the stochastic  gravitational wave  background that generated from cosmological phase transition can be detected by detectors \cite{ch62,ch63,ch64}. Therefore,  this work can provide theoretical support for analyzing the properties of the universe using stochastic gravitational wave background.

\appendix
\setcounter{equation}{0}
\renewcommand\theequation{A.\arabic{equation}}
\section{}
\label{appA}
According to the holographic principle, when a gravitational system absorbs a particle, the area of its horizon and the total energy within it both increase. The minimal change in area $\Delta A$ can be expressed as  \cite{ch66,ch67}:

\vspace{-\baselineskip}
\begin{align}
\label{eqA1}
\Delta A \sim Xm,
\end{align}
with the size  $X$ and mass $m$ of the particle, respectively. In quantum mechanics, the width of the wave packet of a particle is described as the standard deviation of the $X$ distribution, that is the position uncertainty $\Delta x$. Therefore, one has the relationship $X \sim \Delta x$. Moreover, when measuring the position of a particle, the particle's mass must be greater than the uncertainty in its momentum $\Delta p$ \cite{ch68}. Based on the above results, Eq.~(\ref{eqA1}) can be rewritten as

\vspace{-\baselineskip}
\begin{align}
\label{eqA2}
\Delta A \geq \Delta x \Delta p.
\end{align}
The above equation suggests that the smallest increase in area in a gravitational system is constrained by the momentum uncertainty $\Delta p$ and position uncertainty $\Delta x$ of quantum mechanics. By  solving Eq.~(\ref{eq1}), the position uncertainty gives

\vspace{-\baselineskip}
\begin{align}
\label{eqA3}
\Delta p \geq \frac{\hbar }{2}\frac{1}{{\Delta x + 16{\beta  \mathord{\left/ {\vphantom {\beta  {\Delta x}}} \right. \kern-\nulldelimiterspace} {\Delta x}}}}.
\end{align}
For a static spherical gravitational system, the position uncertainty is approximately equal to the radius of the apparent horizon, i.e., $\Delta x \simeq 2 r$ \cite{ch69}, hence, the
minimal change of the area can be expressed as

\vspace{-\baselineskip}
\begin{align}
\label{eqA4}
\Delta A \geq \chi \tilde \hbar \left( {{\beta}} \right),
\end{align}
where $\chi  = 4\ln 2$ and $\tilde \hbar \left( \beta  \right) = {1 \mathord{\left/ {\vphantom {1 {\left[ {2 + \left( {{{32\pi \beta } \mathord{\left/
 {\vphantom {{32\pi \beta } A}} \right. \kern-\nulldelimiterspace} A}} \right)} \right]}}} \right. \kern-\nulldelimiterspace} {\left[ {2 + \left( {{{32\pi \beta } \mathord{\left/ {\vphantom {{32\pi \beta } A}} \right. \kern-\nulldelimiterspace} A}} \right)} \right]}}$ are  the calibration factor and  the effective Planck constant, respectively. Obviously,  when $\beta = 0$, the effective Planck constant reduces to the classical case $\hbar  = {1 \mathord{\left/ {\vphantom {1 2}} \right. \kern-\nulldelimiterspace} 2}$. According to the information theory, the minimal increase of entropy is related to the change in the area reads

 \vspace{-\baselineskip}
 \begin{align}
\label{eqA5}
\frac{{{\text{d}}S}}{{\;{\text{d}}A}} \simeq \frac{{\Delta {S_{\min }}}}{{\Delta {A_{\min }}}} = \frac{1}{{8\tilde \hbar \left( \beta  \right)}}.
\end{align}
In the classical limit, it is well known that he original entropy of a gravitational system is ${S_0} = {A \mathord{\left/ {\vphantom {A 4}} \right. \kern-\nulldelimiterspace} 4}$. However, when accounting for the impact of the GUP, the general expression of entropy needs to be modified to ${S} = {{f\left( A \right)} \mathord{\left/
 {\vphantom {{f\left( A \right)} 4}} \right. \kern-\nulldelimiterspace} 4}$, where ${f\left( A \right)}$ is a function of the area \cite{ch70}

 \vspace{-\baselineskip}
\begin{align}
\label{eqA6}
\frac{{{\text{d}}S}}{{{\text{d}}A}} = \frac{{f'\left( A \right)}}{4}.
\end{align}
Correspondingly, the relationship between entropy and area can be expressed as follows:

\vspace{-\baselineskip}
\begin{align}
\label{eqA7}
\frac{{{\text{d}}S}}{{\;{\text{d}}A}} = \frac{{f'\left( A \right)}}{4},
\end{align}
with $f'\left( A \right) = {{{\text{d}}f\left( A \right)} \mathord{\left/ {\vphantom {{{\text{d}}f\left( A \right)} {{\text{d}}A}}} \right. \kern-\nulldelimiterspace} {{\text{d}}A}}$. By comparing Eq.~(\ref{eqA6}) with Eq.~(\ref{eqA7}), it is found that

\vspace{-\baselineskip}
\begin{align}
\label{eqA8}
f'\left( A \right) = \frac{1}{{2\tilde \hbar \left( \beta  \right)}} = 1 + \frac{{16\pi \beta }}{A}.
\end{align}
Finally, by integrating Eq.~(\ref{eqA7}), the GUP corrected entropy becomes

\vspace{-\baselineskip}
\begin{align}
\label{eqA9}
{S_{{\text{GUP}}}} = \int {\frac{{f'(A)}}{{4G}}}{\text{d}}A = \frac{A}{4} + 4\pi \beta \ln A.
\end{align}
In the limit $\beta\rightarrow 0$, one yields the  original entropy  $S_0$.

\section{}
\label{appB}
\renewcommand\theequation{B.\arabic{equation}}
Therefore, it is necessary to derive the modified Friedmann equations in the framework of Eq.~(\ref{eq1+}). In a homogeneous and isotropic spacetime, the FRW universe can be described by the following line element

\vspace{-\baselineskip}
\begin{align}
\label{eqx1}
{\text{d}}{s^2} = {h_{\mu \nu }}{\text{d}}{x^\mu }{\text{d}}{x^\nu } + R^2 \left( {{\text{d}}{\theta ^2} + {{\sin }^2}\theta {\text{d}}{\varphi ^2}} \right),
\end{align}
where ${x^\mu } = \left( {t,r} \right)$ with $\mu  = \nu = 0$, $1$, $R= r  a\left(t\right)$ with the scale factor $a\left(t\right)$, and ${h_{\mu \nu }} = \operatorname{diag} \left[ { - 1,{a^2}} \right]$  is the two-dimensional metric with the spatial curvature constant  $k$. According to the condition  ${h^{\mu \nu }}\left( {{\partial _u}R} \right)\left( {{\partial _\nu }R} \right) = 0$, the dynamical apparent of FRW universe  reads  ${R_A} = a\left( t \right)r = {H^{ - 1}}$, with the Hubble parameter $H$. When considering the matter of the FRW universe as a perfect fluid, its energy-momentum tensor can be expressed as

\vspace{-\baselineskip}
\begin{align}
\label{eqx2}
{T_{\mu \nu }} = (\rho  + p){u_\mu }{u_\nu } + p{g_{\mu \nu }},
\end{align}
where  $\rho$ is the energy density, $p$ is the pressure, $u_\mu$ is the four velocity of the fluid, and $g_{\mu\nu}$ is the spacetime metric of FRW universe. Based on the law of conservation of energy-momentum $T_{;\nu }^{\mu \nu } = 0$,  the continuity equation is given by $\dot \rho  + 3H\left( {\rho  + p} \right) = 0$. According  to Refs.~\cite{ch18,ch19},  the matter within the apparent horizon satisfy the first law of thermodynamics:

\vspace{-\baselineskip}
\begin{align}
\label{eqx3}
{\text{d}}E =  - T{\text{d}}S + W{\text{d}}V,
\end{align}
where $E= \rho V$ is the total of all the matter energy in the apparent horizon, $V = {{4\pi {R_A^3}} \mathord{\left/ {\vphantom {{4\pi {R_A^3}} 3}} \right. \kern-\nulldelimiterspace} 3}$ denotes the 3-dimensional sphere's volume, and $W = {{\left( {\rho  - p} \right)} \mathord{\left/ {\vphantom {{\left( {\rho  - p} \right)} 2}} \right. \kern-\nulldelimiterspace} 2}$ is the density of work. According to the continuity equation, the differential form of energy in the form of

\vspace{-\baselineskip}
\begin{align}
\label{eqx4}
{\text{d}}E = \rho {\text{d}}V + V{\text{d}}\rho  = 4\pi R_A^2\rho {\text{d}}{R_A} + \frac{{4\pi R_A^3}}{3}{\text{d}}\rho .
\end{align}
In the FRW universe, the surface gravity on the apparent horizon can be defined as $\kappa  =  - {{\left( {1 - {{{{\dot R}_A}} \mathord{\left/ {\vphantom {{{{\dot R}_A}} 2}} \right. \kern-\nulldelimiterspace} 2}} \right)} \mathord{\left/ {\vphantom {{\left( {1 - {{{{\dot R}_A}} \mathord{\left/ {\vphantom {{{{\dot R}_A}} 2}} \right.
 \kern-\nulldelimiterspace} 2}} \right)} {{R_A}}}} \right. \kern-\nulldelimiterspace} {{R_A}}}$, and the temperature of the apparent horizon is ${{T = \kappa } \mathord{\left/ {\vphantom {{T = \kappa } {2\pi }}} \right. \kern-\nulldelimiterspace} {2\pi }}$.  Now, by using the expression of entropy and the temperature of the apparent horizon,  the equation for the first law of thermodynamics on the right hand side can be rewritten as

\vspace{-\baselineskip}
\begin{subequations}
\label{eqx5}
\begin{align}
T{\text{d}}S & =  - \frac{1}{{2\pi {R_A}}}\left( {1 - \frac{{{{\dot R}_A}}}{{2H {R_A}}}} \right)\frac{{f'(A)}}{4}{\text{d}A},
\\
W{\text{d}}V  & = 2\pi {R_A^2}\left( {\rho  - p} \right){\text{d}}R_A,
\end{align}
\end{subequations}
Substituting Eq.~(\ref{eqx5}) and the dynamical apparent horizon of the FRW universe $R$ into Eq.~(\ref{eqA8}), the first Friedmann equation takes the form as follows

\vspace{-\baselineskip}
\begin{align}
\label{eqx6}
 - \frac{{8\pi }}{3}\left( {\rho  + p} \right) = \dot Hf'\left( A \right)
 \end{align}
Next, substituting the continuity equation into the Eq.~(\ref{eqx6}), and then integrating, one has the second Friedmann equation

\vspace{-\baselineskip}
\begin{align}
\label{eqx7}
\frac{8}{3}\pi\rho  =  - 4\pi \int {f'\left( A \right)} \frac{{{\text{d}}A}}{{{A^2}}},
 \end{align}
In the framework of Eq.~(\ref{eq2}), the GUP corrected Friedmann equations read

\vspace{-\baselineskip}
\begin{subequations}
\label{eqx8}
\begin{align}
- 4\pi \left( {\rho  + p} \right) & = \left( {\dot H - \frac{k}{{{a^2}}}} \right)\left( {1 + \frac{{16\pi \beta }}{A}} \right),
\\
\frac{8}{3}\pi \rho &  = \frac{{4\pi \left( {A + 8\pi \beta } \right)}}{{{A^2}}} + \mathcal{C},
 \end{align}
 \end{subequations}
where $\mathcal{C}$ is an integration constant, it can be determined by considering the boundary conditions in the vacuum energy (dark energy) dominated epoch. In this epoch, as the area of the cosmic apparent horizon becomes $A \rightarrow \infty$, the energy density becomes $\rho = \rho_{\text{vacuum}}= \Lambda$. Therefore, one gets  $\mathcal{C} = 3{{8\pi \Lambda } \mathord{\left/ {\vphantom {{8\pi \Lambda } 3}} \right. \kern-\nulldelimiterspace} 3}$ \cite{ch65}. When considering that $A = 4\pi {R_A^2} = {{4\pi } \mathord{\left/ {\vphantom {{4\pi } {{H^2}}}} \right. \kern-\nulldelimiterspace} {{H^2}}}$, the GUP corrected Friedmann equations  are~\cite{ch54}

\vspace{-\baselineskip}
\begin{subequations}
\label{eqx9}
\begin{align}
- 4\pi \left( {\rho  + p} \right) & = \dot H\left( {1 + 4\beta {H^2}} \right),
\\
\frac{8}{3}\pi \left( {\rho  - \Lambda } \right) & = {H^2}\left( {1 + 2\beta {H^2} } \right).
  \end{align}
 \end{subequations}
In order to highlight the role of effects of QG, one needs to study the early universe era, the observed cosmological constant is enough tiny. For the sake of simplicity, we assume that cosmological constant $\Lambda$ can be ignored. Therefore, the GUP corrected Friedmann equations  are~\cite{ch54}

\vspace{-\baselineskip}
\begin{subequations}
\label{eqx10}
\begin{align}
 - 4\pi (\rho  + p) &= \dot H\left( {1 + 4\beta {H^2}} \right),
 \\
\frac{8}{3}\pi \rho & = {H^2}\left( {1 + 2\beta {H^2}} \right).
\end{align}
\end{subequations}

\section{}
\label{appC}
\renewcommand\theequation{C.\arabic{equation}}
In Refs.~\cite{ch19,ch22}, the unified first law is given by

\vspace{-\baselineskip}
\begin{align}
\label{eqa1}
\text{d}\tilde E = \tilde A\tilde \Psi  + \tilde W \text{d}\tilde V,
\end{align}
where $\tilde A = 4\pi {R^2}$ is the area of a sphere with  radius $R$, and $\tilde W =  - {{{h^{ab}}{T_{ab}}} \mathord{\left/ {\vphantom {{{h^{ab}}{T_{ab}}} 2}} \right.  \kern-\nulldelimiterspace} 2}$ is the work density of the matter fields. The energy flux  $\tilde \Psi$  is defined as

\vspace{-\baselineskip}
\begin{align}
\label{eqa2}
\tilde \Psi  = {\tilde \Psi _a}{\text{d}}{x^a} = \left( {T_a^b{\partial _b}R + \tilde W{\partial _a}R} \right){\text{d}}{x^a},
\end{align}
with the energy-momentum tensor of the perfect fluid ${T_{ab}} = (\rho  + p){u_a}{u_b} + p{g_{ab}}$. According to above expressions, the  energy supply vector can be expressed as follows

\vspace{-\baselineskip}
\begin{align}
\label{eqa3}
{\tilde \Psi _a} = \left[ { - \frac{1}{2}\left( {\rho  + p} \right)HR,\frac{1}{2}\left( {\rho  + p} \right)a} \right].
\end{align}
Substituting Eq.~(\ref{eqa3}) into Eq.~(\ref{eqa2}), one has

\vspace{-\baselineskip}
\begin{align}
\label{eqa4}
\tilde A\tilde \Psi  & = \tilde A{{\tilde \Psi }_a}\text{d}{x^a} = 2\pi {R^2}\left( {\rho  - p} \right)\left( { - HR \text{d}t + a \text{d}r} \right)
\nonumber \\
& =  - \frac{{\tilde A}}{2}\left( {\frac{{1 + 4{H^2}\beta }}{{4\pi }}} \right)\dot H\left( {\text{d}R - 2HR \text{d}t} \right).
\end{align}
At the apparent horizon of the FRW universe $R_A$, Eq.~(\ref{eqa4}) is rewritten as

\vspace{-\baselineskip}
\begin{align}
\label{eqa5}
{\left. {\tilde A\tilde \Psi } \right|_{R = {R_A}}}  &= \frac{A}{2}\left( {1 + \frac{{4\beta }}{{R_A^2}}} \right)\frac{{{{\dot R}_A}}}{{4\pi R_A^2}}\left( {\text{d}{R_A} - 2\text{d}t} \right) \nonumber \\
&= \frac{A}{2}\left( {1 + \frac{{4\beta }}{{R_A^2}}} \right)\frac{{{{\dot R}_A}}}{{4\pi R_A^2}}\text{d}{R_A} \!-\! \frac{A}{2}\left( {1\! +\! \frac{{4\beta }}{{R_A^2}}} \right)\frac{{\text{d}{R_A}}}{{2\pi R_A^2}}
\nonumber \\
&= \frac{{{{\dot R}_A}}}{2}\left( {1 + \frac{{4\beta }}{{R_A^2}}} \right)\text{d}{R_A} \!-\! \left( {1 + \frac{{4\beta }}{{R_A^2}}} \right)\text{d}{R_A}
\nonumber \\
&=  - \frac{1}{{2\pi {R_A}}}\left( {1 - \frac{{{{\dot R}_A}}}{2}} \right)\left( {2\pi {R_A} + \frac{{8\pi \beta }}{{{R_A}}}} \right)\text{d}{R_A}.
\end{align}
For the second term of the right hand side of Eq.~(\ref{eqa1}) at the apparent horizon of the FRW universe becomes

\vspace{-\baselineskip}
\begin{align}
\label{eqa6}
{\left. {\tilde W} \text{d} {\tilde V} \right|_{R = {R_A}}} = \frac{{4\pi R_A^3}}{6}\left( {\rho  - p} \right)= W \text{d} {V}.
\end{align}
Now, according to Eq.~(\ref{eqa5}) and Eq.~(\ref{eqa6}), the unified first law takes the form

\vspace{-\baselineskip}
\begin{align}
\label{eqa7}
\text{d}E =  - T\text{d}S + W\text{d}V.
\end{align}
where ${\left. {\tilde E} \right|_{R = {R_A}}} = E$. Obviously, the above equation is the  thermodynamic first law.

\end{document}